\documentclass[twocolumn]{aastex63}

\usepackage[utf8]{inputenc} 
\usepackage{amsmath, amsbsy}
\usepackage{tikz}
\usetikzlibrary{bayesnet} 
\usepackage{enumitem} 

\shorttitle{5 nucleosynthetic channels from LAMOST}
\shortauthors{Wheeler et al.}

\newcommand{\gaia}{\emph{Gaia}}
\newcommand{\thecannon} {\emph{The Cannon}}
\newcommand{\lamost}{LAMOST}
\newcommand{\galah}{GALAH}
\newcommand{\apogee}{APOGEE}
\newcommand{\insitu}{\emph{in-situ}}
\newcommand{\ts}{\textsuperscript}
\newcommand{\feh}{\mathrm{[Fe/H]}}
\newcommand{\tonfe}[1]{{[#1/Fe]}}

\newcommand{\teff}{T_\mathrm{eff}}
\newcommand{\logg}{\log(g)}
\newcommand{\vmic}{v_\mathrm{mic}}
\newcommand{\onfe}[1]{{[#1/\mathrm{Fe}]}}
\newcommand{\T}{T}
\newcommand{\vtheta}{\ensuremath{\boldsymbol\theta}}
\newcommand{\vell}{\ensuremath{\boldsymbol\ell}}
\newcommand{\veta}{\ensuremath{\boldsymbol\eta}}
\newcommand{\kpc}{\mathrm{kpc}}
\newcommand{\kms}{\mathrm{km~s^{-1}}}
\newcommand{\kpckms}{\ensuremath{\mathrm{kpc~km~s^{-1}}}}

\newcommand\reviewer[1]{{#1}}

\begin{document}
\title{Abundances in the Milky Way across five nucleosynthetic channels from 4 million LAMOST stars}

\author[0000-0001-7339-5136]{Adam Wheeler}
\affiliation{Department of Astronomy, Columbia University, Pupin Physics Laboratories, New York, NY 10027, USA}

\author{Melissa Ness}
\affiliation{Department of Astronomy, Columbia University, Pupin Physics Laboratories, New York, NY 10027, USA}
\affiliation{Center for Computational Astrophysics, Flatiron Institute, 162 Fifth Avenue, New York, NY 10010, USA}

\author{Sven Buder}
\affiliation{Max Planck Institute for Astronomy (MPIA), Koenigstuhl 17, 69117 Heidelberg, Germany}
\affiliation{Research School of Astronomy \& Astrophysics, Australian National University, Canberra, ACT 2611, Australia}
\affiliation{ARC Centre of Excellence for All Sky Astrophysics in 3 Dimensions (ASTRO 3D), Canberra, ACT 2611, Australia}

\author{Joss Bland-Hawthorn}
\affiliation{Sydney Institute for Astronomy, School of Physics, A28, The University of Sydney, NSW 2006, Australia}
\affiliation{ARC Centre of Excellence for All Sky Astrophysics in 3 Dimensions (ASTRO 3D), Canberra, ACT 2611, Australia}

\author{Gayandhi De Silva}
\affiliation{Australian Astronomical Optics, Macquarie University, 105 Delhi Rd, North Ryde, NSW 2113, Australia}

\author{Michael Hayden}
\affiliation{Sydney Institute for Astronomy, School of Physics, A28, The University of Sydney, NSW 2006, Australia}
\affiliation{ARC Centre of Excellence for All Sky Astrophysics in 3 Dimensions (ASTRO 3D), Canberra, ACT 2611, Australia}

\author{Janez Kos}
\affiliation{University of Ljubljana, Faculty of Mathematics and Physics, Jadranska 19, 1000 Ljubljana, Slovenia}

\author{Geraint F. Lewis}
\affiliation{Sydney Institute for Astronomy, School of Physics, A28, The University of Sydney, NSW 2006, Australia}

\author{Sarah Martell}
\affiliation{School of Physics, UNSW, Sydney, NSW 2052, Australia}
\affiliation{ARC Centre of Excellence for All Sky Astrophysics in 3 Dimensions (ASTRO 3D), Canberra, ACT 2611, Australia}

\author{Sanjib Sharma}
\affiliation{Sydney Institute for Astronomy, School of Physics, A28, The University of Sydney, NSW 2006, Australia}
\affiliation{ARC Centre of Excellence for All Sky Astrophysics in 3 Dimensions (ASTRO 3D), Canberra, ACT 2611, Australia}

\author{Jeffrey D. Simpson}
\affiliation{School of Physics, UNSW, Sydney, NSW 2052, Australia}

\author{D. B. Zucker}
\affiliation{Department of Physics and Astronomy, WW7 2.705, Macquarie University, NSW 2109 Australia}
\affiliation{Macquarie University Research Centre for Astronomy, Astrophysics \& Astrophotonics, Sydney, NSW 2109, Australia
}

\author{Tomaž Zwitter}
\affiliation{University of Ljubljana, Faculty of Mathematics and Physics, Jadranska 19, 1000 Ljubljana, Slovenia}

\correspondingauthor{Adam Wheeler}
\email{a.wheeler@columbia.edu}

\begin{abstract}
Large stellar surveys are revealing the chemodynamical structure of the Galaxy across a vast spatial extent.
However, the  many millions of low-resolution spectra observed to date are yet to be fully exploited.
We employ \thecannon, a data-driven approach for estimating \reviewer{chemical abundances},
to obtain detailed abundances from low-resolution (R = 1800) LAMOST spectra, using the GALAH survey as our reference.
We deliver five (for dwarfs) or six (for giants) estimated abundances representing five different nucleosynthetic channels, for 3.9 million stars, to a precision of 0.05 - 0.23 dex.
Using wide binary pairs, we demonstrate that our abundance estimates provide chemical discriminating power beyond metallicity alone.
We show the coverage of our catalogue with radial, azimuthal and dynamical abundance maps, and examine the neutron capture abundances across the disk and halo, which indicate different origins for the \insitu\ and accreted halo populations.
LAMOST has near-complete \emph{Gaia} coverage and provides an unprecedented perspective on chemistry across the Milky Way.
\end{abstract}


\section{Introduction}

Large stellar surveys such as \gaia\ \citep{gaia:16}, \apogee\ \citep{majewski:17, holtzman:18, garciaperez:16}, \galah\ \citep{desilva:15, Martell:2017}, \emph{Gaia}-ESO \citep{gilmore:12}, RAVE \citep{steinmetz:06}, \lamost\ \citep{Deng:2012, Zhao:2012} and SEGUE \citep{yanny:09} are providing the data to empirically characterize the Milky Way disk and infer the primary drivers of its formation and evolution \reviewer{\citep{Freeman_Bland-Hawthorn_2002, Bland-Hawthorn_Gerhard_2016}}.
 
Detailed \reviewer{chemical abundances} are one of the primary measurements made from stellar spectra.
Their determination is a primary motivation for medium- and high-resolution spectroscopic surveys for several reasons: 
they provide effective chemical fingerprints of stars , link directly to the environment in which they were born \citep[e.g.][]{Krumholz2019}, and describe the chemical diversity of the disk (e.g. \citealp{weinberg:19}) and the chemical pathways of enrichment \citep[e.g.][]{rybizki:17}.  Combined with stellar  kinematics, abundances are core to the pursuit of Galactic archaeology. 

Conventionally, detailed abundances have been derived from medium- and high- resolution stellar spectra (e.g. \apogee: $R=22,500$, \galah: $R=28,000$, and \reviewer{RAVE : $R=7500$, \emph{Gaia}-ESO: at least $R = 20,000$})
Until recently, the inferences from low-resolution spectra, such as \lamost\ and SEGUE, were typically limited to stellar parameters and $\alpha$-enhancements ($\teff$, $\logg$, [Fe/H], [$\alpha$/Fe]) (e.g. \citealp{lee:11}).
\citet{ting:18_oxygen} have shown that oxygen abundances can be inferred from spectra in wavelength regions containing no atomic oxygen lines through the features of species in the CNO atomic-molecular network.
Indirectly inferred abundances also have a long empirical history.
The Ca II triplet, for example, is an often used metallicity index (\citealp{Armandroff:88}; see \citealp{Vasquez:15} for a recent calibration).
To date, with the exception of \citet{xiang:19}, the efforts to extract individual abundances from \lamost\ have largely focused on a few elements, namely an integrated $\alpha$-element abundance and the elements C and N, which are particularly important, as these elements can indicate age (e.g. \citealt{li:16, xiang:17, ho:17a, ho:17b, zhang:19}).

In this work, we use a data-driven approach to label low-resolution \lamost\ spectra with several abundances.
\lamost\ is one of the largest stellar surveys to date, with over 5 $\times$ 10$^{6}$ publicly available spectra, at $R=1800$.
The survey has extensive coverage of the Milky Way's disk, halo and, in particular, the outer disk, the detailed chemodynamics of which are largely unexplored.
Specifically, we employ \thecannon{} \citep{ness:15}, a model characterized in large part by its simplicity, to derive individual abundances from \lamost.
Other data-driven methods include \emph{The Payne} \citep{ting:2019_payne}, which,  like \thecannon\, works by explicitly modelling spectra as a function of labels (stellar parameters and abundances), and that of \citet{leung:18}, which uses a \reviewer{convolutional neural network} to estimate labels directly from spectra without explicit inference.
\citet{xiang:19} recently released a catalog of 16 abundances (C, N, O, Na, Mg, Al, Si, Ca, Ti, Cr, Mn, Fe, Co, Ni, Cu, and Ba) for \lamost\ DR 5 using a neural-net-based model calibrated by both labelled spectra (using overlap between \lamost\ and both \apogee\ and \galah) and physical modelling.
This work has many common aspects with our own, but is different in detail.
\reviewer{Both calibrate flexible spectral models (a shallow neural network, in the case of \citealp{xiang:19}) with labels from high-resolution surveys, but \citet{xiang:19} also employ gradients of \emph{ab-initio} models.
An advantage of using model gradients is that physical expectations are incorporated into the label derivation.
Our approach, however, prioritises the data alone in specifying the model, which can be advantageous when physical models are lacking.
Differences between the catalogues for those elements trained using the \galah\ labels will help reveal the biases of each approach. }

Our approach requires reference objects, stars with high-quality spectra and precise labels (stellar parameters and abundances), that are representative of the survey objects.
They are used to calibrate a model that produces synthetic spectra from stellar labels.
This model is then used to estimate labels for the full set of survey stars, in our case, the \lamost\ catalog.
Both the \apogee\ and \galah\ surveys have stars in common with \lamost\ which can serve as possible reference objects.
\apogee\  provides higher precision abundance measurements than \galah, which enables, for example, the clear disambiguation of the the low- and high- $\alpha$ sequences, as seen in the radial maps of \citet{hayden:15} and \citet{nidever:14}.
However, the dimensionality of the abundance space measured by \apogee\ is low \citep{ness:18, pricejones:18, ness:19} (although note that weak lines of neutron capture elements have been identified in this region \citep{Cunha2017, Hasselquist2016}).
\galah, on the other hand, provides abundance measurements across a more extensive set of nucleosynthetic channels, including the neutron-capture ($r$ and $s$) processes.
The neutron-capture element enhancements have been previously explored only through boutique analyses of small samples of stars observed at high resolution (e.g. \citealp{bensby:14, spina:18}) and in the solar neighbourhood, to which \galah\ is largely confined (e.g. \citealp{buder:19, Schonrich:19}). 
\galah\ also provides abundances for main-sequence stars, allowing us to extend our modelling to that regime.

We want to explore the promise of the largest number of element abundance families as possible, so we took the roughly 10,000 stars in common between \galah\ and \lamost\ to build a model using the \lamost\ spectra and \galah\ stellar parameters and abundances.
While the \galah\ labels are less precise than those from \apogee\, and thus yield less precise \lamost\ labels, the \lamost\ catalog is large enough to enable very precise mean estimates of abundances on a population basis (e.g \citealp{ness:19, blancato:19}).
Using \galah\ as a source for our input labels allows us to propagate \emph{r}-process and \emph{s}-process abundances to the outer disk and halo. 

In deriving a set of individual abundances for \lamost, this work complements the \lamost\ catalogue, which provides stellar parameters and bulk metallicity (a term used interchangeably with [Fe/H] in this work) only.
We deliver inferred abundances for elements from five nucleosynthetic families:
\emph{light} elements, which are dispersed by asymptotic giant branch (AGB) stars and core-collapse supernovae (CCSN), 
and whose atmospheric abundances can change due to dredge-up;
$\alpha$-elements, which are dispersed primarily by CCSN;
\emph{iron-peak} elements, which are dispersed by both CCSN and type Ia supernovae (SNIa);
\emph{odd-Z} elements, which are dispersed by both CCSN and SNIa and expected to display similar trends to the $\alpha$ elements;
$s$-process elements, which are thought to be produced and dispersed in AGB stars;
and $r$-process elements, which are produced in extremely neutron-rich environments. 
It is not clear at present whether neutron-star mergers are the primary site of the $r$-process, or if other sites make appreciable contributions \reviewer{(e.g. \citealp{Arnould:07, Cote:18, Hansen_Holmbeck_Beers_Placco_Roederer_Frebel_Sakari_Simon_Thompson_2018, Siegel2019, Sakari_Roederer_Placco_Beers_Ezzeddine_Frebel_Hansen_Sneden_Cowan_Wallerstein_et-al_2019, Sakari_Placco_Farrell_Roederer_Wallerstein_Beers_Ezzeddine_Frebel_Hansen_Holmbeck_et_al_2018})}.
For each star, we deliver five (for dwarfs) or six (for giants) abundances of O (light), Eu ($r$-process), mean $\alpha$, Sc (iron-peak), mean $s$-process, Mg ($\alpha$), Al (odd $Z$), Mn (iron-peak), and Ba ($s$-process).
Having derived these abundances, we demonstrate the scientific value of multi-element abundances of large numbers of stars.
We do this using pairs of stars across the disk and halo, examining the abundance similarity of wide binaries, that have been identified by their kinematics alone.
We also map the chemodynamical abundance structure of the disk and halo, making links to signatures of  evolution such as radial migration and Galaxy assembly. 

In \S2 we describe the \galah\ and \lamost\ data and the quality cuts we applied.  
\S3 provides a brief overview of \thecannon.
In \S4 we discuss model checks and evaluate the error of our label estimates.
\S5 discusses our public catalog and key scientific results, and \S6 discusses their implications.

\section{data} 
\begin{figure}
    \includegraphics[width=0.45\textwidth]{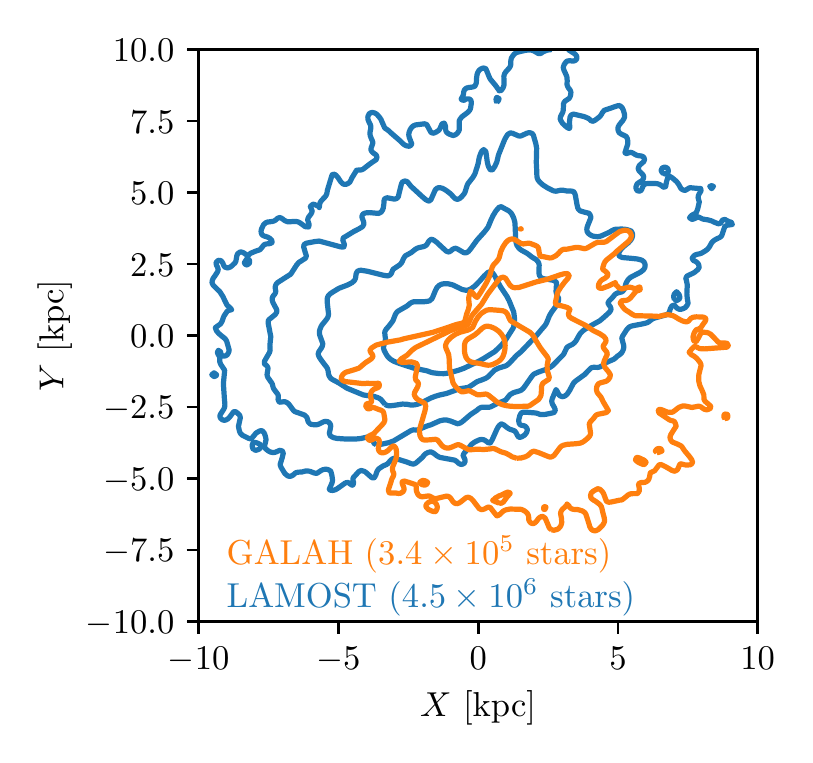}
    \includegraphics[width=0.45\textwidth]{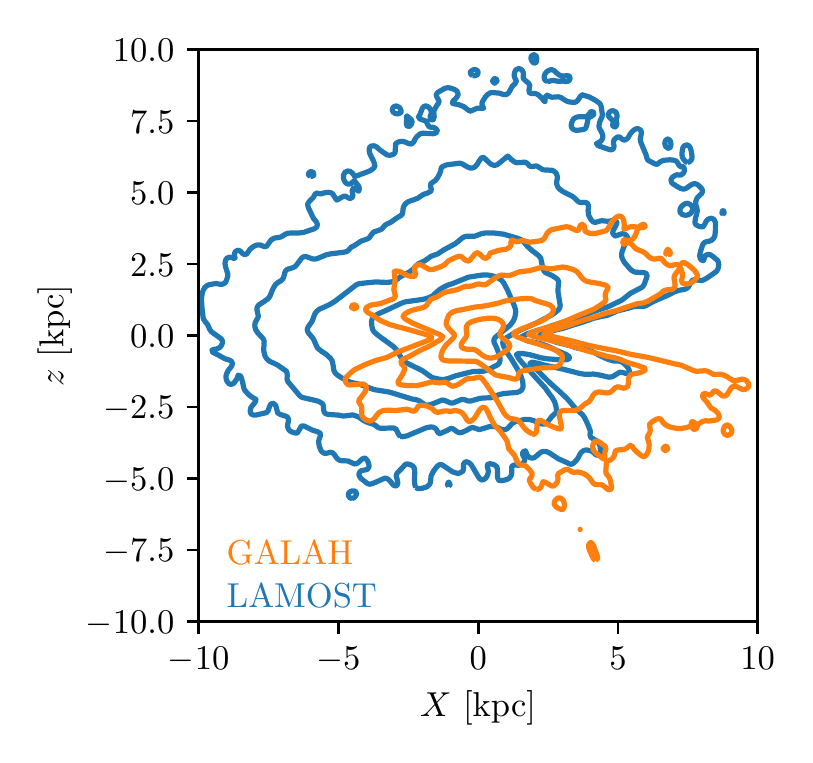}
    \caption{Face on (top) and edge on (bottom) Contours in surface density ($5\times 10^{4,3,2,1}$ kpc$^{-2}$) shown in heliocentric Galactic coordinates for \galah\ and \lamost, which probes much farther into the outer disk and halo.
    The Galactic center is at $X = 8~\kpc, Y=0$.}
    \label{fig:GVLfootprint}
\end{figure}

\begin{figure*}
    \includegraphics[width=\textwidth]{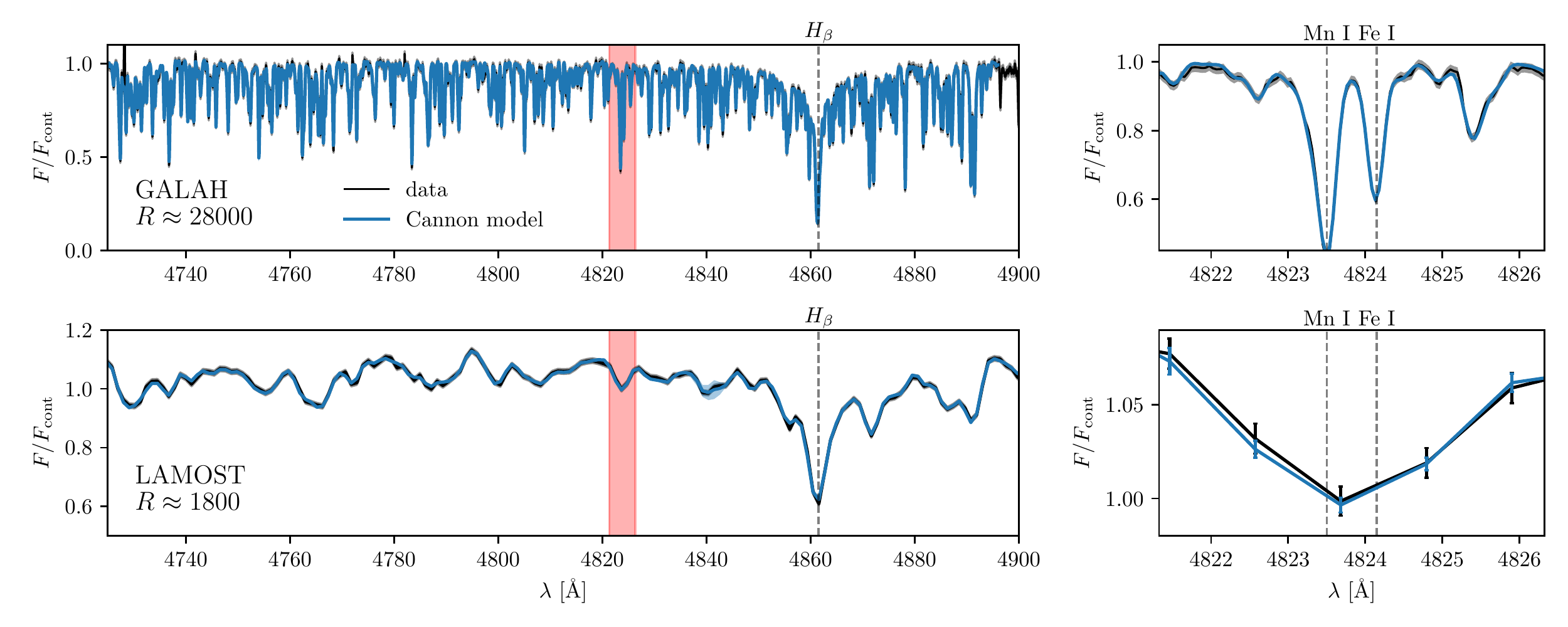}
    \caption{A comparison of the \galah\ and \lamost\ spectra for the same star, 2MASS 00010184+0407201, \emph{Gaia} DR2 2740354684364096000.
    The top panels show part of the star's \galah\ spectrum ($\mathrm{S/N} = 65$) and \emph{Cannon} model, while the bottom two show the same for LAMOST ($\mathrm{S/N} = 179$).
    On the left, note the large $H_\beta$ line and surrounding features, on the right, observe the fit around a known Mn feature (highlighted in red on the left).
    Both the measured spectra and \emph{Cannon} models are shown with their 1-$\sigma$ errors, shown with error bars in the right-hand panels and a shaded region in the left-hand panels.
    }
    \label{fig:GVLspectra}
\end{figure*}

Our data comprise the R=1800 DR 4 v2 \lamost\ spectra, the R=28,000 DR 2.1 \galah\ spectra and stellar parameter and abundance labels \citep{buder:18}, as well as the \gaia\ proper motion and parallax measurements for our stars.
From \galah\ we use $\teff$, $\logg$, $\vmic$, and [Fe/H], along with abundances with respect to Fe, of O, Si, Ca, Ti, Eu, Sc, Y, Mg, Al, Mn, and Ba.
Figure \ref{fig:GVLfootprint} shows the Galactic footprints of \galah\ and \lamost.
A portion of each survey's spectrum for a typical training set star is shown in Figure \ref{fig:GVLspectra}.

\subsection{Quality cuts and data cleaning}
One of the formal assumptions of \thecannon\ is that the training labels are known exactly, so constructing a high-fidelity training set is crucial.
To build our training set, we first determined the set of stars in common between \galah\ and \lamost.
We performed a $1''$ sky match between \galah\ DR 2.1 and \lamost\ DR 4 v2 to identify these reference object candidates, of which there were roughly ten thousand.
We then removed all stars from the potential training set with signal to noise ratio (S/N) less than $30$ in either the LAMOST $z$ band (\texttt{snrz}) or the \galah\ blue channel (\texttt{snr\_c1}). 
We also removed any star for which \texttt{chi2\_cannon} (a column in the \galah\ catalog, not a product of our analysis) was greater than $4$, which indicates that the best fit spectral model is a poor fit to the whole spectrum, and any star for which \texttt{flag\_cannon} was nonzero, which can indicate a variety of problems with abundance determination.
These cuts removed roughly half of the stars from consideration.

We found that cutting on the reported \galah\ label errors did not improve our performance against the validation set.  
To further exclude low-quality measurements from our training set, we therefore generated and evaluated the fit of the best-fit \emph{Cannon} model spectrum for each reference stellar spectrum, for every element in the \galah\ catalog.
The \galah\ pipeline uses separate \emph{Cannon} models for each elemental abundance in order to restrict each model to the wavelengths of unblended lines.
Each model has different best-fit parameters, which we were not able to retrieve. 
They are, however, within the errors of the mean reported stellar parameters for each star \citep{buder:18}.
For the stellar parameter labels, we used the values in the \galah\ DR2.1 catalog\footnote{available at \url{https://docs.datacentral.org.au/galah/}.}, along with $A_K$ values calculated with the Rayleigh-Jeans color excess (RJCE) method \citep{majewski:11} applied to ALLWISE \citep{wright:10, mainzner:11} and 2MASS \citep{skrutskie:06} broadband photometry, as was done for the \galah\ models. 
We calculated $\chi^2$ between the best-fit \galah\ model and the observed \galah\ spectrum for every star in our training set in the region of the strongest lines of each element (the \texttt{chi2\_cannon} flag pertains to the global fit).
Appendix \ref{sec:windows} lists the wavelength regions used, which are the same windows used in the \galah\ pipeline.
The distribution of $\chi^2$ values for some elements peaked lower than expected from nominal measurement error alone by a factor of 2-3, meaning that a cut on some multiple of $\chi^2/\mathrm{dof}$ was not theoretically justified.
We removed all stars with $\chi^2$ values above the 85\ts{th} percentile, for any of its abundances. This led to a significant improvement in our cross-validation results, as discussed in our methods, on the order of 15\%-40\% percent). 
Using the 75\ts{th} percentile,  as a more conservative cut, gave us no improvement in cross-validation tests.
These cuts leave 1722 stars in the training set.
We do not exclude stars flagged in \galah\ based on \texttt{flag\_x\_fe} because we performed our own per-abundance $\chi^2$ cut and because removing stars where the \galah\ model may be extrapolating reduces the size of our training set too drastically.
We emphasize however that \thecannon\ is likely to extrapolate well (in a well understood way using the simple polynomial model we employ) for many abundances.

\subsection{Dwarf and giant models} 

\begin{figure}
    \centering
    \includegraphics[width=0.45\textwidth]{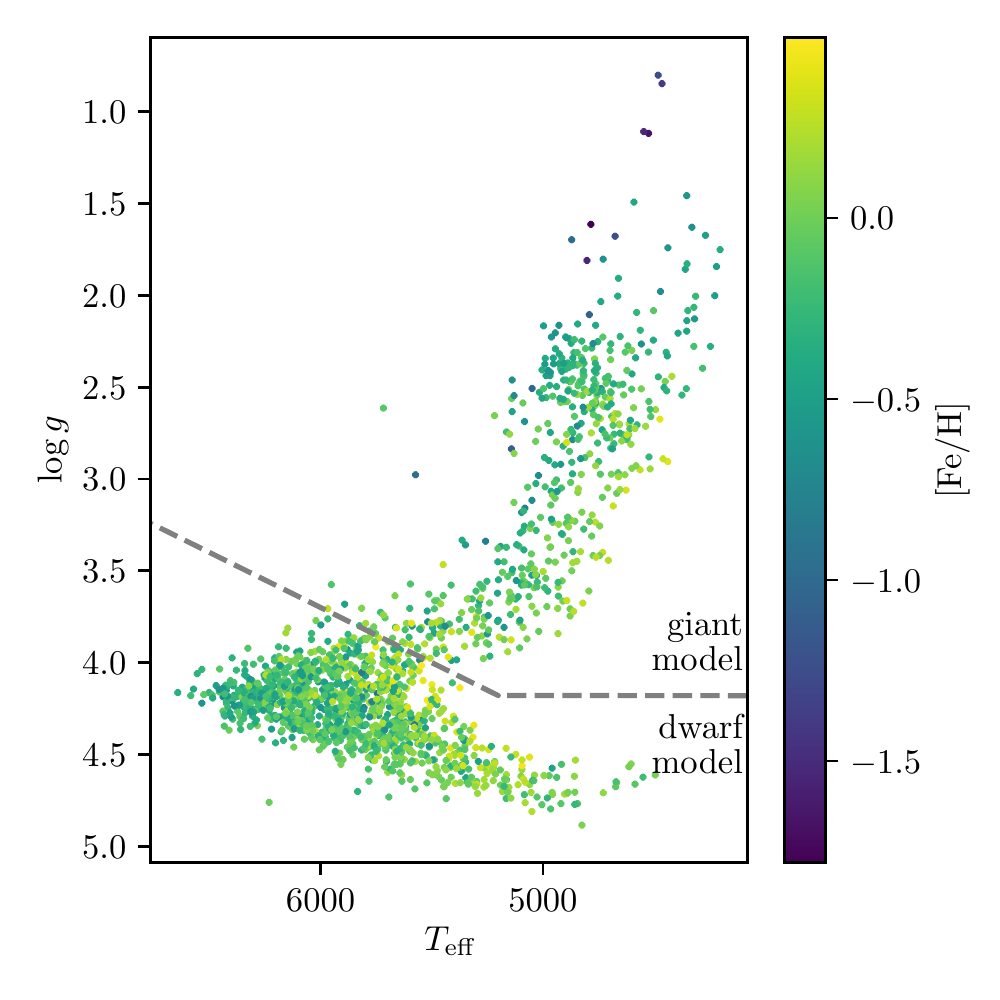}
    \caption{
    Our 1722 training stars in the \lamost\ Kiel diagram space colored by (\galah) [Fe/H]. Since there are more metal-poor stars in the giant training set, our giant model is unbiased down to lower metallicity.
    }
    \label{fig:trainingHR}
\end{figure}

\begin{figure*}
    \centering
    \includegraphics{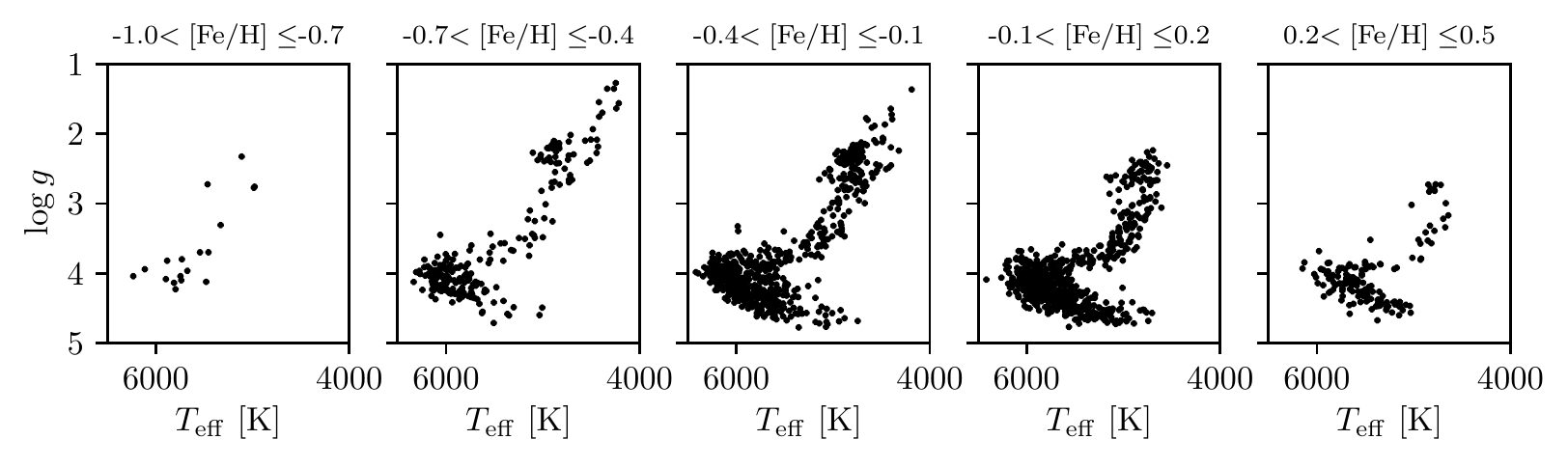}
    \caption{\reviewer{Our 1722 training stars across the (\lamost) Kiel diagram in bins of metallicity. For the extreme ends of our metallicity range, the lack of training set coverage may bias our estimated labels.}}
    \label{fig:metallicitybins}
\end{figure*}

After the quality cuts described in Section 2.1, we were left with a training set that spans \reviewer{the Kiel diagram and metallicity (Figures \ref{fig:trainingHR},  \ref{fig:metallicitybins}).}
We modeled giants and dwarfs separately, with the division between models given in terms of \lamost\ $\log g$ and $\teff$ by 
\begin{equation}\label{eq:split}
    \log g = \begin{cases}
        4.18 & \teff < 5200 \mathrm{K}\\ 
    (-6 \times 10^{-4}) \teff/\mathrm{K} + 7.3 & \teff \geq 5200 \mathrm{K} .
    \end{cases}
\end{equation} 

The split gives us 532 giants and 1190 dwarfs as our reference objects.  
\reviewer{The values in Equation (\ref{eq:split}) to separate dwarfs and giants are somewhat arbitrary.
We find that the performance of each model is not not sensitive to these precise values.}
We decided which elements to infer for each model by balancing our ability to recover each abundance in cross-validation (\S\ref{sec:modelchecking}) with the objective of having several elements across nucleosynthetic channel.
For both models, we include $\teff$, $\log g$, $\vmic$, [Fe/H], \tonfe{O}, and \tonfe{Eu} as labels.
For the dwarfs, we also used iron-relative abundances of: error-weighted mean $\alpha$ (from Mg, Si, Ca, and Ti), Sc, and error-weighted mean s-process (from Ba and Y), while for the giants we also used Mg ($\alpha$), Al (odd-$Z$), Mn (iron-peak), and Ba ($s$-process).  
Training a model without $\alpha$ or Mg \reviewer{yields} a systematic offset in inferred neutron-capture abundances, likely because the model will exploit correlations between these nucleosynthetic families if they are not controlled for.

We only used mean abundances in the same nucleosynthetic family if they appeared strongly correlated in the training set.
We tried using dereddened \gaia\ \emph{G} band magnitude instead of $\log g$, which would allow us to apply a prior at test time, but we found that this did not improve our results in practice.
The model had trouble predicting extinction, partially because our training sets do not include any high-extinction stars.
Including extinction as a label did not improve our ability to predict any of the abundances, so we opted not to.

\begin{figure}
    \centering
    \includegraphics{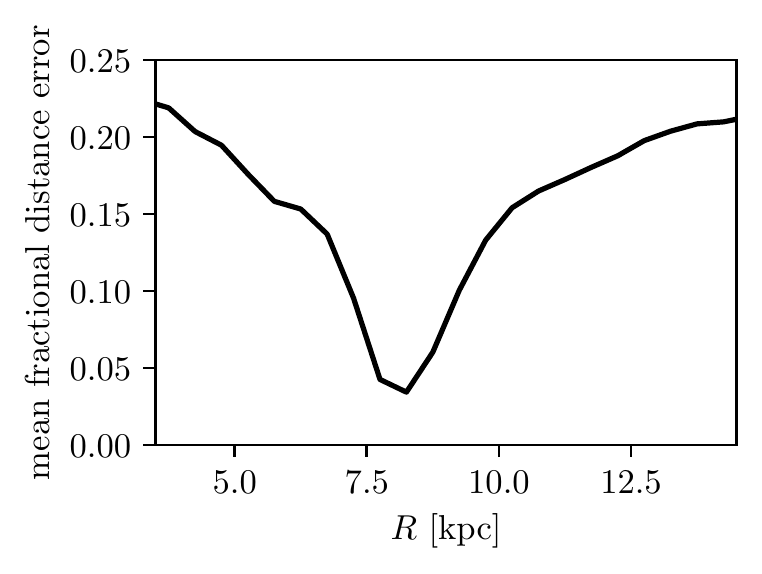}
    \caption{Mean fractional distance error as a function of Galactocentric radius. Distance errors blur chemical maps of the Milky Way far from the solar annulus.}
    \label{fig:disterrs}
\end{figure}

For subsequent analysis, we cross-matched with \emph{Gaia} by taking the source within 1 arcsecond of the LAMOST star with the lowest \emph{G}-band magnitude.
Throughout this paper, we use the parallactic distance estimates from \citet{bailerjones:18}, \reviewer{which makes use of a prior incorporating the expected galactic spatial distribution.
Other distance catalogs (e.g. \citealp{Anders:2019}) will have different biases for stars with uncertain distances (e.g. those far from the Sun), so quantitative results derived from our catalog will be conditioned upon the assumptions of \citet{bailerjones:18}.
Our results below are largely qualitative, and unlikely to be strongly dependant on choice of distance catalog.
Figure \ref{fig:disterrs} shows the mean fractional distance error as a function of Galactocentric radius, $R$.
Distance errors blur our maps of chemistry across the Galaxy (Section \ref{sec:maps}), particularly far from the solar annulus, where they reach 20\%.
}

\section{Model}  

The following is a brief description of \thecannon\ \citep[see][for a more extended discussion.]{ness:15}.
For this work we build a \textsc{julia}-based implementation of \thecannon, which is documented and available at \hyperlink{https://github.com/ajwheeler/TheCannon.jl}{this URL}\footnote{\url{github.com/ajwheeler/TheCannon.jl}} and via the \textsc{julia} package manager.
The source code uses the same nomenclature as the description here, and allows for optional masking of labels (self-consistent training with the model constrained so that each label is only ``on'' at specified wavelengths).

For each star, $n$, we take the flux value in the spectral pixel with wavelength $\lambda$ to be  $F_{n \lambda}$ and its (Gaussian, independent) measurement uncertainty to be $\sigma_{n \lambda}$.
To prepare the spectra for \thecannon\ we first redshift-corrected the spectra using the \texttt{z} value provided in the LAMOST data table 
and interpolated each star to a common wavelength grid.
We then continuum-normalized the spectra by dividing out the continuum, approximated by smoothing the spectra with a Gaussian kernel with a 50 \AA\  standard deviation, truncated at 150 \AA\ from the center, in the same manner as \citet{ho:17a, ho:17b}. 
This normalized flux is then near unity in the absence of emission or absorption features.
For each reference star, we also define $\vell_n$ be the vector containing its physical parameters and abundances (its \emph{labels}).
These are the quantities we ultimately wish to infer for the rest of the LAMOST spectra, at test time. 

Our labels vector $\vell_n$ for each reference stars is: 
\begin{equation}
    \vell_n = \begin{bmatrix}
    \teff & \log(g) & \vmic & \mathrm{[Fe/H]} & \mathrm{[X}_1\mathrm{/Fe]} & \dots  & \mathrm{[X}_N\mathrm{/Fe]} \end{bmatrix}^\T
\end{equation}
where $\mathrm{X}_1, \dots, \mathrm{X}_N$ are the elements whose abundances we wish to determine.
It is good practice for both numerical stability and model flexibility to express all labels in units such that they are distributed around zero and have similar magnitudes.
We do this by subtracting from each label its (training set) mean and dividing it by its (training set) dispersion.
This transformation is then undone after the inference has taken place.

In numerous published uses of \thecannon\ (including this one), the flux in each pixel is described by a 2\ts{nd} degree polynomial of the elements of the label vector whose coefficients, $\vtheta$, are determined by a training set of spectra for which, ideally, both accurate and precise labels are available.
 For a given spectral pixel and star, we then have our spectral flux, $F$, for our $n$ reference objects at each wavelength, $\lambda$ defined as: 

\begin{align*}
    F_{n \lambda} &= \theta^{0}_\lambda  &&\text{(constant term)}\\[5pt]
                  &+ \theta^{\teff}_\lambda\teff + \dots + \theta^{X_N }_\lambda\onfe{X_N} &&\text{(linear terms)}\\[5pt]
                  &+ \theta^{\teff^2}_\lambda\teff^2 + \dots + \theta^{X_N^2}_\lambda(\onfe{X_N})^2 && \text{(squared terms)} \\[5pt]
                  &+ \theta^{\teff \log(g)}_\lambda\teff\log(g) + \dots  \\
                  &\quad+ \theta^{X_NX_{N-1}}_\lambda \onfe{X_N} \onfe{X_{N-1}}&&\text{(cross-terms)} \\[5pt]
    &+ \mathrm{error}.
\end{align*} \label{eq:poly}

To specify an error model, we can write the above as a likelihood function
\begin{equation}
    F_{n \lambda} | \vell_n, \vtheta_\lambda, s_\lambda \sim \mathcal{N}(\veta(\vell_n) \cdot \vtheta_\lambda, \sigma_{n \lambda}^2 + s_\lambda^2)
\end{equation}
where $\mathcal{N}$ is the normal distribution, $s_\lambda$ is model uncertainty (either inherent stochasticity or physics that hasn't been captured by the model) at wavelength $\lambda$, 
$\vtheta_\lambda$ is the vector of coefficients describing how the flux at $\lambda$ varies with label value, 
and $\veta$, the quadratic expansion (called the vectorizing function in \citet{casey:16}), maps from labels to every 0\ts{th}, 1\ts{st}, and 2\ts{nd} order combination of components of the label vector,  
\begin{equation}
    \veta(\vell_n) = \begin{bmatrix} 1 & T_\mathrm{eff} & \dots & \mathrm{[X}_N\mathrm{/Fe]} & T_\mathrm{eff}^2 & T_\mathrm{eff}\log(g) & \dots \end{bmatrix}^\T.
\end{equation}
If $\vell_n$ is a vector of length $N$, $\veta(\vell_n)$ is a vector of length $(N^2 + 3N)/2$.
A more flexible model could be constructed by replacing $\veta$ to an expansion with higher order terms, or to other combinations of labels.
However, quadratic models have been shown to be sufficient in practice \reviewer{\citep{ness:15, ness:16, ness:19, ho:17a, ho:17b}.
In fact, a linear model is often all that is needed \citep{Birky_Hogg_Mann_Burgasser_2020, Hogg_Eilers_Rix_2018}. }
The combinatoric increase in model parameters that would be necessary for a higher order polynomial is undesirable.
\begin{figure}
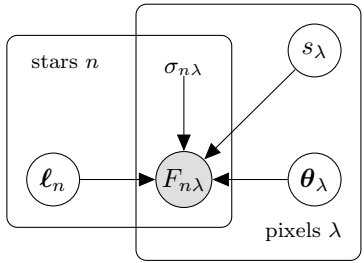

  \centering
  \tikz{
      \node[latent] (l) {$\vell_n$} ; %
      \node[obs, right=of l] (F) {$F_{n \lambda}$} ; %
      \node[const, above=of F] (sigma) {$\sigma_{n \lambda}$} ; %
      \node[latent, right=of F] (theta) {$\vtheta_\lambda$} ; %
      \node[latent, above=of theta] (s) {$s_\lambda$} ; %

      \plate[inner sep=0.25cm] {pixels} {(F) (sigma) (theta) (s)} {pixels $\lambda$};
      \tikzset{plate caption/.append style={right=2pt of #1.north west}}
      \plate[inner sep=0.25cm] {stars} {(l) (F) (sigma)} {stars $n$}; 

      \edge {sigma} {F} ; %
      \edge {l} {F} ; %
      \edge {s} {F} ; %
      \edge {theta} {F} ; %
  }
  \caption{\thecannon\ likelihood as a probabilistic
      graphical model.  During training, the latent variables in the ``stars''
      panel ($\vell_n$) are fixed using stars for which labels are known,  in
      our case from \galah.  When inferring stellar labels, the latent variables
      in the ``pixels'' panel ($\vtheta_\lambda$ and $s_\lambda$) are fixed to their point
  estimates from training and the maximum likelihood estimates for all $\vell_n$ are
  calculated.}
  \label{fig:PGM}
\end{figure}

Figure~\ref{fig:PGM} shows the likelihood function as a probabilistic graphical model, which depicts the relationships between observed and latent quantities.
Ideally, the full joint distribution over training data and output labels would be sampled from directly (with e.g. Markov-chain Monte-Carlo), but such an approach is not computationally feasible.
Instead the problem is divided into a training step, in which a point estimate of each $\vtheta_\lambda$ and $s_\lambda$ is estimated from the training set, and an inference step, in which the labels of each star are estimated.

During the training step, each $\vtheta_\lambda$ and $s_\lambda$ is jointly fixed to its maximum-likelihood estimate (MLE), given the labeled spectra in the training set.
For fixed $s_\lambda$, the model is linear with fixed Gaussian error in $\vtheta_\lambda$, so its MLE, $\widehat{\vtheta}|_{s_\lambda}$, can be calculated analytically.
Finding $\widehat{s_\lambda}$ is then a matter of numerically maximizing the log-likelihood,
\begin{equation}
\begin{split}
    \log \mathcal{L}(s_\lambda) = - \sum_n \frac{1}{2} \biggl ( \frac{(\veta(\ell_n)\cdot \widehat{\vtheta_\lambda}|_{s_\lambda} - F_{n\lambda})^2}{\sigma_{n \lambda}^2 + s_\lambda^2} \\
        + \ln(\sigma_{n \lambda}^2 + s_\lambda^2) \biggr ) + \mathrm{const}, 
\end{split}
\end{equation}
in one dimension.
During the inference step, the MLE of $\vell_n$ is calculated with $s_\lambda$ and $\vtheta_\lambda$ fixed to their point estimates.
There is no trick to get us out of multivariate optimization here, since $\veta$ is nonlinear.

\section{Model Evaluation} \label{sec:modelchecking} 
\begin{figure*}
    \includegraphics[width=\linewidth]{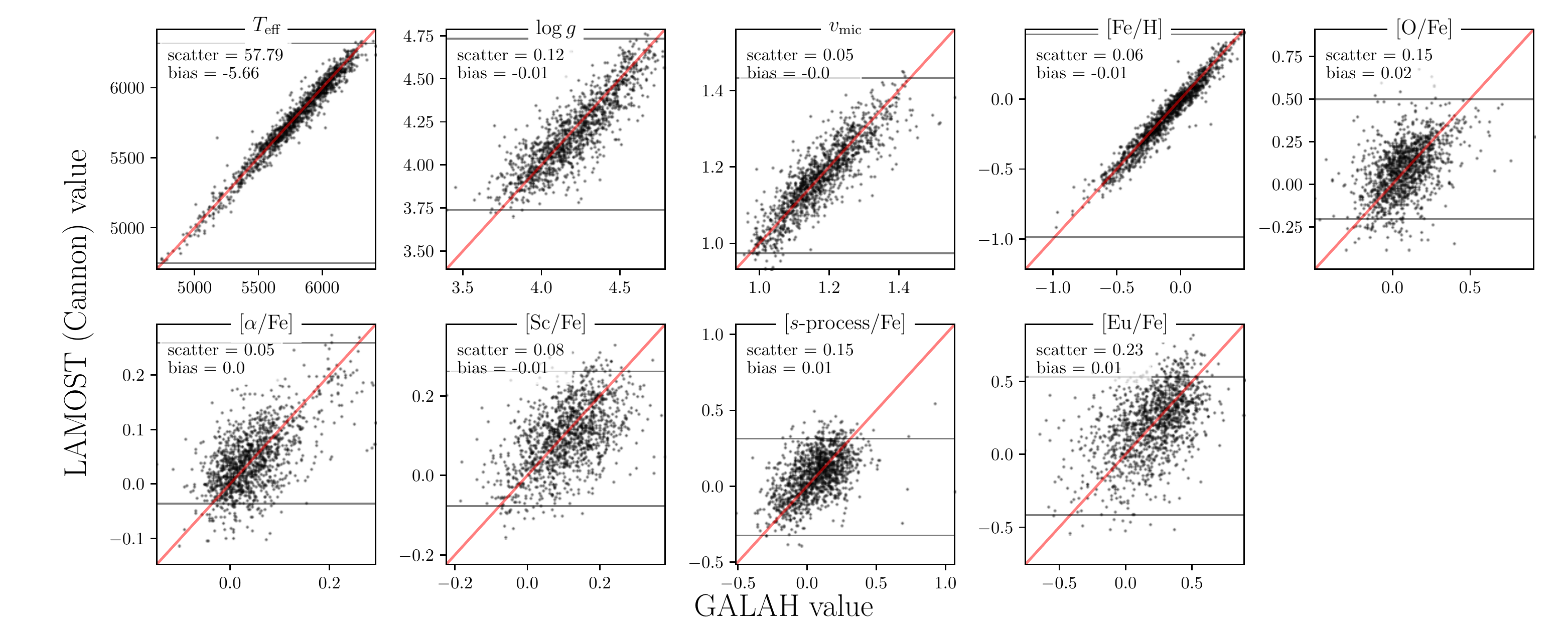}
    \caption{Cross-validation recovery of training set labels for the dwarf model.
    Horizontal lines show boundaries beyond which the model is biased or unprobed by the training set.}
    \label{fig:dwarfsCV}
\end{figure*}

\begin{figure*}
    \includegraphics[width=\textwidth]{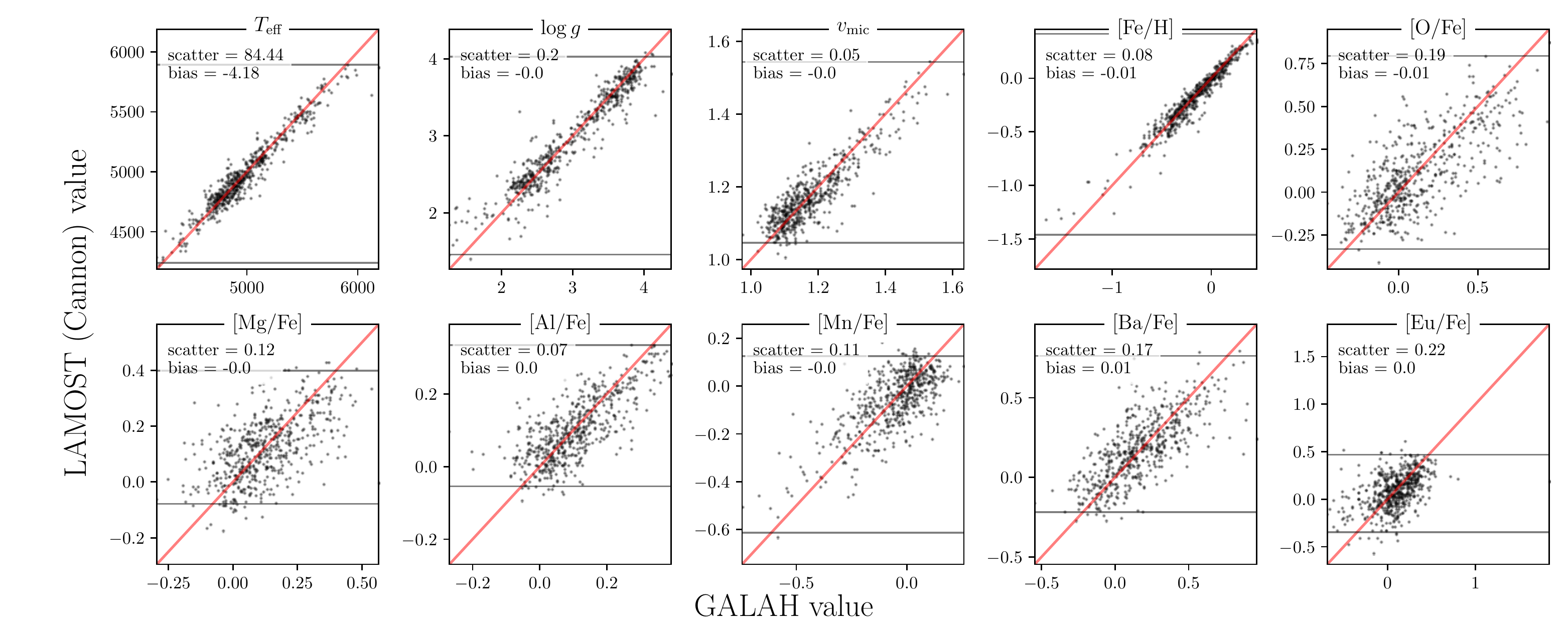}
    \caption{Analogous to Figure \ref{fig:dwarfsCV}.
    Cross-validation recovery of training set labels for the giant model.
    Horizontal lines show boundaries beyond which the model fails to extrapolate. 
    While the scatter in [Fe/H] is higher at lower values, the model appears to be nearly unbiased down to $\mathrm{[Fe/H]} \approx -1.5$.}
    \label{fig:giantsCV}
\end{figure*}

We use 12-fold cross-validation (CV) in order to verify that the model is able to recover stellar labels.
We partition the reference objects into twelve random subsets, then predict the labels of each subset using the other eleven as training data.
This gives us a prediction for each reference star that has not leveraged its \galah\ labels.
Figures \ref{fig:dwarfsCV} and \ref{fig:giantsCV}  show CV performance for the giants and dwarfs respectively, along with the scatter, bias, and correlation coefficient for each label.  
Our CV-assessed abundance precision ranges from 0.05 to 0.23 dex for dwarfs and 0.07 to 0.22 dex for giants.
Examination of the labels inferred for spectra from repeat observation of the same star show differences consistent with CV-precision.

\begin{figure}
    \centering
    \includegraphics[width=0.45\textwidth]{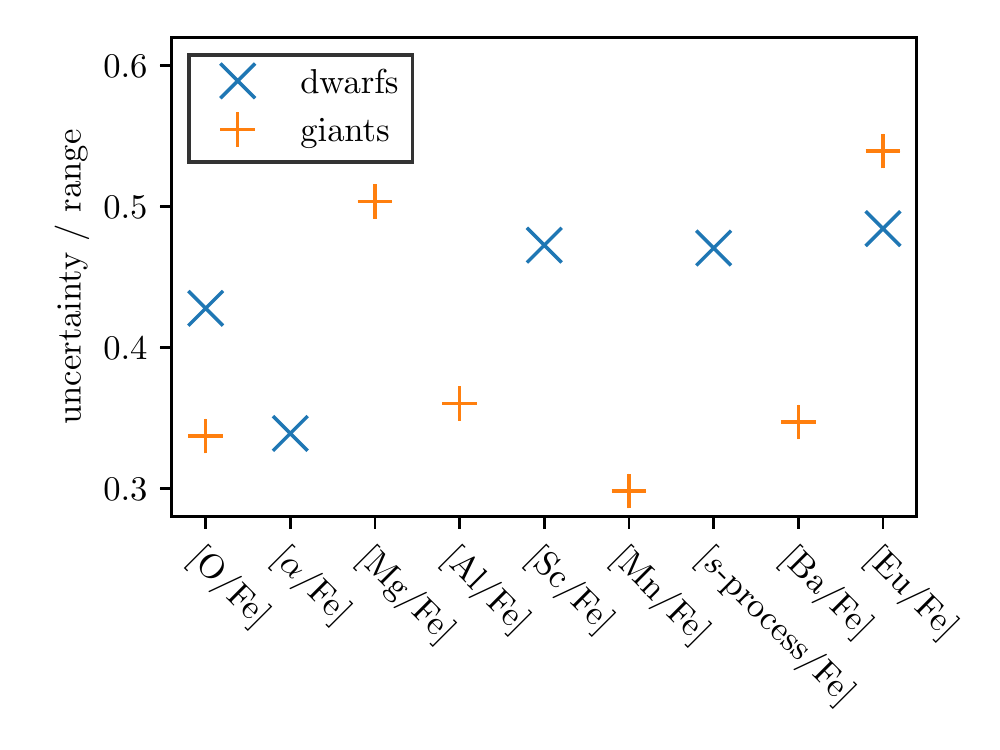}
    \caption{
    The precision of each of our abundances relative to the range over which the model is approximately unbiased.
    We generally infer abundance ratios with precision at the 30-50\% level. Smaller values indicate higher relative precision of that abundance and presumably higher discriminating power between stars. 
    }
    \label{fig:reluncert}
\end{figure}

We also use CV to identify the thresholds beyond which our model is highly biased or unprobed by the training data.
We say that the model is in this regime when it under- or over-predicts the label being considered 90\% of the time in CV.
The specific calculation is as follows:
For a given label, $l$ (e.g. $l = \teff$), we approximate $p(l_\mathrm{true}, l_\mathrm{inferred})$ with a kernel-density estimate (KDE) with bandwidth chosen by Silverman's rule \citep{silverman:86}, then use this approximate distribution to find the values of $l_\mathrm{inferred}$ at which $p(l_\mathrm{true} | l_\mathrm{inferred})$ excludes $l_\mathrm{inferred}$ at the 90\% level.
These boundaries are shown as horizontal lines in Figures \ref{fig:dwarfsCV}, and \ref{fig:giantsCV}. Stars that fall beyond these boundaries are flagged in our catalog.
Figure \ref{fig:reluncert} shows the precision (twice the scatter in Figures \ref{fig:dwarfsCV} and \ref{fig:giantsCV}) of each of our abundances relative to the range over which the model is roughly unbiased.
This quantity is often what is relevant when comparing the labels of different stars, rather than characterizing a single star. 

\subsection{Model interpretability}

\thecannon\ is simple enough that its parameters are open to direct interpretation.
This sets it apart from more complex modeling approaches such as, e.g. neural networks.
 
It is clear, for example, that our model learns $\teff$ in large part from the Balmer series, as this is where $\theta_{\teff}$ becomes large.
By examining model scatter, $s_\lambda$ as a function of wavelength, we can tell that our model is less precise in the regions of CN bands, at the beginning of the spectral region of \lamost.
In some wavelength regions, $s_\lambda$ drops to 0, likely because continuum-normalization introduces small correlations between nearby pixels that are not accounted for by the model.
It is also apparent that the model is leveraging the whole spectrum to predict abundances, rather than strong lines only.
We performed tests by isolating only regions where individual abundance features are present in the spectra, forced did not allow the coefficients to vary from zero at training time outside of these regions.
This is approach is similar to the way in which \galah\ determined their abundance labels, using so called abundance windows.
For the \lamost\ spectra, this approach of using windows fails to recover abundance ratios in cross-validation.
Section \ref{sec:cat} discusses some implications of this fact, and our scatter and linear coefficients are plotted as a function of wavelength in Appendix \ref{app:coeffs}.

\section{Results} 

\begin{figure*}
    \centering
            \includegraphics[width=\textwidth]{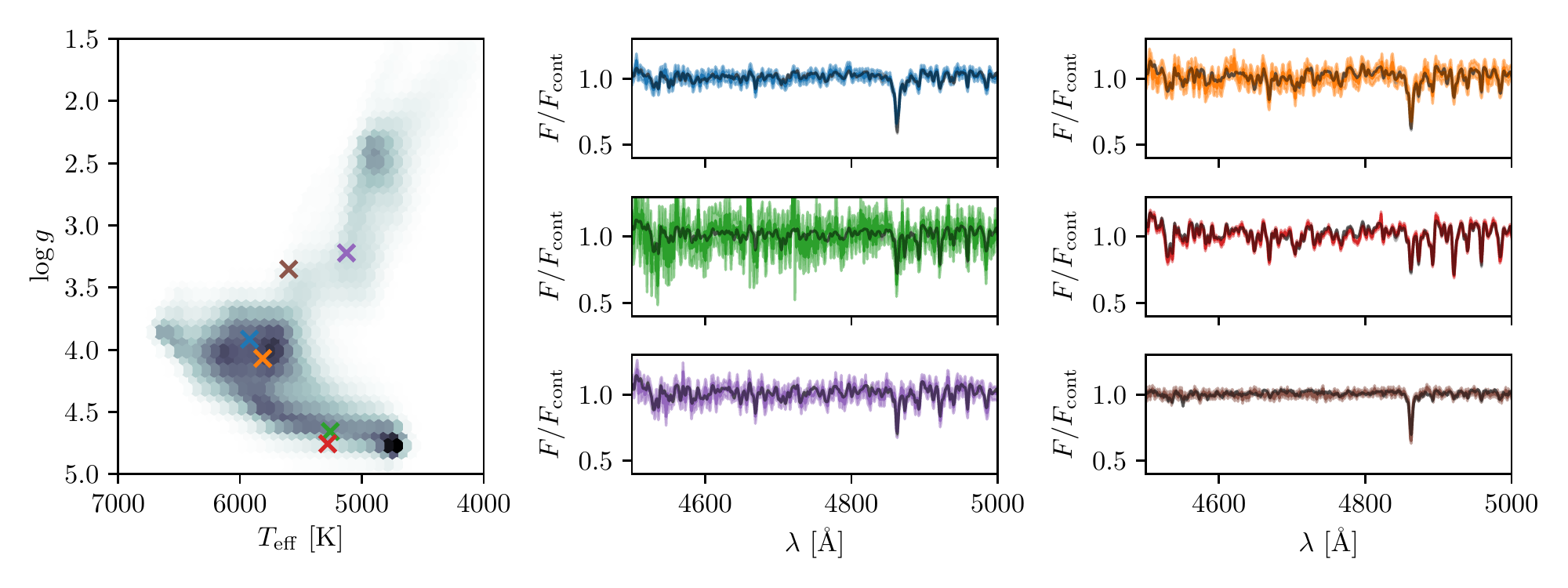}
    \caption{A portion of the best-fit model spectra (black) and real data (colored), both with 1-$\sigma$ uncertainties, for 6 randomly-chosen stars in our catalog. 
    Though simple, the model is flexible enough to fit the data across the Kiel diagram.}
    \label{fig:catspectra}
\end{figure*}

\subsection{Catalog} \label{sec:cat}

We produce a catalog of stellar parameters and individual abundances for 4,541,883 observations of 3,744,284 stars across the Kiel diagram (Figure \ref{fig:catspectra}), which is available \href{https://doi.org/10.7910/DVN/5VWKMC}{at this URL}, 
along with our model coefficients and training set.
We combine observations of the same star by reporting ($z$-band) $S/N$-weighted averages of their labels.
Along with our inferred stellar parameters and abundances, we provide the \lamost\ and \gaia\ identifiers for each star, as well as its Galactic position, radial velocity and estimated actions.
We also provide windowed and whole-spectrum $\chi^2$ values and flags to tell when the model is extrapolating.
For each star, we calculated approximate actions with \texttt{galpy} \citep{bovy:15} using the St\"{a}ckel fudge \citep{binney:12, bovy:13} and \texttt{MWPotential2014}, applied using \gaia\ distances \citep{bailerjones:18} and proper motions, and LAMOST radial velocities (RVs).
\reviewer{While \gaia\ achieves a better RV precision (see Figure \ref{fig:RVprec}), \gaia\ RVs are only available for approximately one fifth of our catalog.}
We assumed that the Sun sits at $X = 8~\mathrm{kpc}$, $z = 0.025~\mathrm{kpc}$ \citep{juric:08} and is moving with $v_X = 11.1~\kms$, $v_Y = -232.24~\mathrm{km~s^{-1}}$, $v_z = 7.25~\kms$ \citep{schoenrich:09}.
Table \ref{tab:schema} provides the full catalog schema.
\reviewer{To understand the effect of RV and distance uncertainty on our estimated actions, we sampled their values from their error distribution and calculated Galactocentric coordinates and actions for each iteration, performed 20 times.
The median uncertainties for $J_R$, $J_\phi$, and $J_z$, respectively were $5~\kpckms$, $21~\kpckms$, and $1~\kpckms$.
Appendix \ref{sec:kinematicerrors}  explores these errors in more detail.
}

\begin{deluxetable*}{lllp{10cm}}
\tablecaption{ Catalog schema.
Here, \texttt{x} and X stand for each of the chemical symbols for the elements whose abundances being estimated.
Our Galactic coordinate system is right-handed.
We also make available the table of per-observation labels.
}
\label{tab:schema}
 \tablehead{
 \colhead{column name} & \colhead{type} & \colhead{unit} & \colhead{description}
 }
 \startdata 
\texttt{source\_id} & integer & & \gaia\ DR2 source id  \\
\texttt{designation} & string & & \lamost\ unique star identifier  \\
\texttt{giantmodel} & boolean & & true if labels were estimated with giant model  \\
\texttt{teff} & float & K & $\teff$ \\
\texttt{logg} & float  &  & $\logg$ \\
\texttt{vmic} & float & & $\vmic$\\
\texttt{kiel\_extrap} & boolean & & true if $\teff$ or $\logg$ (the axes of the Keil diagram) are in regime where model fails to extrapolate for any observation\\
\texttt{chi2} & float & & whole-spectrum $\chi^2$\\
\texttt{fe\_h} & float & & [Fe/H] \\
\texttt{fe\_h\_extrap} & boolean & & true if [Fe/H] value is in regime where model fails to extrapolate for any observation\\
\texttt{x\_fe} & float & & [X/Fe] \\
\texttt{chi2\_x\_fe} & float & & $\chi^2$ calculated in windows around strong lines of X\\
\texttt{x\_fe\_extrap} & boolean & & true if [X/Fe] value is in regime where model fails to extrapolate for any observation \\
\texttt{snrz} & float & & LAMOST $z$-band $S/N$\\
\texttt{ra} & float & deg & right-ascension\\
\texttt{dec} & float & deg & declination\\
\texttt{R} & float & kpc &  $R$, in Galactic cylindrical coordinates\\
\texttt{phi} & float & rad & $\phi$, in Galactic cylindrical coordinates\\
\texttt{z} & float & kpc & $z$, in Galactic cylindrical coordinates\\
\texttt{vR} & float &  $\kms$ &  $R$-velocity\\
\texttt{vT} & float & $\kms$ & $\phi$-velocity\\
\texttt{vz} & float & $\kms$ & $z$-velocity\\
\texttt{JR} & float & $\kpckms$ & radial action\\
\texttt{Jphi} & float & $\kpckms$ & angular momentum\\
\texttt{Jz} & float & $\kpckms$ & vertical action\\
 \enddata
\end{deluxetable*}

Here we highlight several caveats to the use of these data:
\begin{itemize}[leftmargin=*]
    \item We allow our model to take leverage of the full information content of the spectrum.
    It therefore not only learns from the most fundamental features of each label, but from correlated features (as we can identify using our model coefficients, which is an advantage of a simple interpretable model).
    Examination of the model's coefficients reveals that the whole spectrum is leveraged in order to predict each abundance.
    Our CV tests show that our model works. 
    It performs well with no hyperparameter tuning, and our analysis of wide binaries in the solar neighborhood \citep{el-badry:19} is indicative of the additional discriminating power beyond an overall metallicity these abundances provide (see Section \ref{sec:widebinaries}).
    However, the abundances are not being measured directly.
    The fidelity of our predicted labels relies on our reference objects (confined to the solar neighborhood) being representative of the survey data.
    For this reason, our reported abundances may be more accurate for disk stars than halo stars.
    In order to identify many case where the model fails to generalize from the training set, we provide $\chi^2$ values calculated across the whole spectrum and individually in narrow windows centered on strong lines, for each element.
    If the best-fit spectrum is a poor fit around known features of a given element, it is likely highly enriched or depleted in that element.
    In fact, this approach is a good way to find such stars with anomalous abundance patterns.
    Indeed, \citet{casey:19}, \citet{kemp:18}, and \citet{Norfolk:19} have used the departure of a basic stellar parameter model generated with \thecannon, from the spectra, to find LAMOST stars that are enhanced in Li, K, and Ba and Sr, respectively.
    
    \item A caveat which is general to data-driven methods is that the model will not necessarily extrapolate correctly outside the parameter space spanned by the training set \reviewer{(see Figure \ref{fig:metallicitybins})}.
    We provide flags to indicate when individual abundances are in the regime where they may be incorrectly extrapolated, as well as a flag indicating when $\teff$, $\logg$, $\vmic$, or [Fe/H] may be incorrectly extrapolated (Table \ref{tab:schema}).
    We determine when our model is extrapolating as described previously, in Section \ref{sec:modelchecking}.
    
    \item While error estimates for each abundance ratio are desirable, producing accurate ones would be prohibitively costly with our current inference infrastructure.
    We advise the user to utilize our CV-assessed error and caution them to be aware that treating our abundance estimates as homoscedastic is a necessary compromise.

    \begin{figure}
        \centering
        \includegraphics[width=0.5\textwidth]{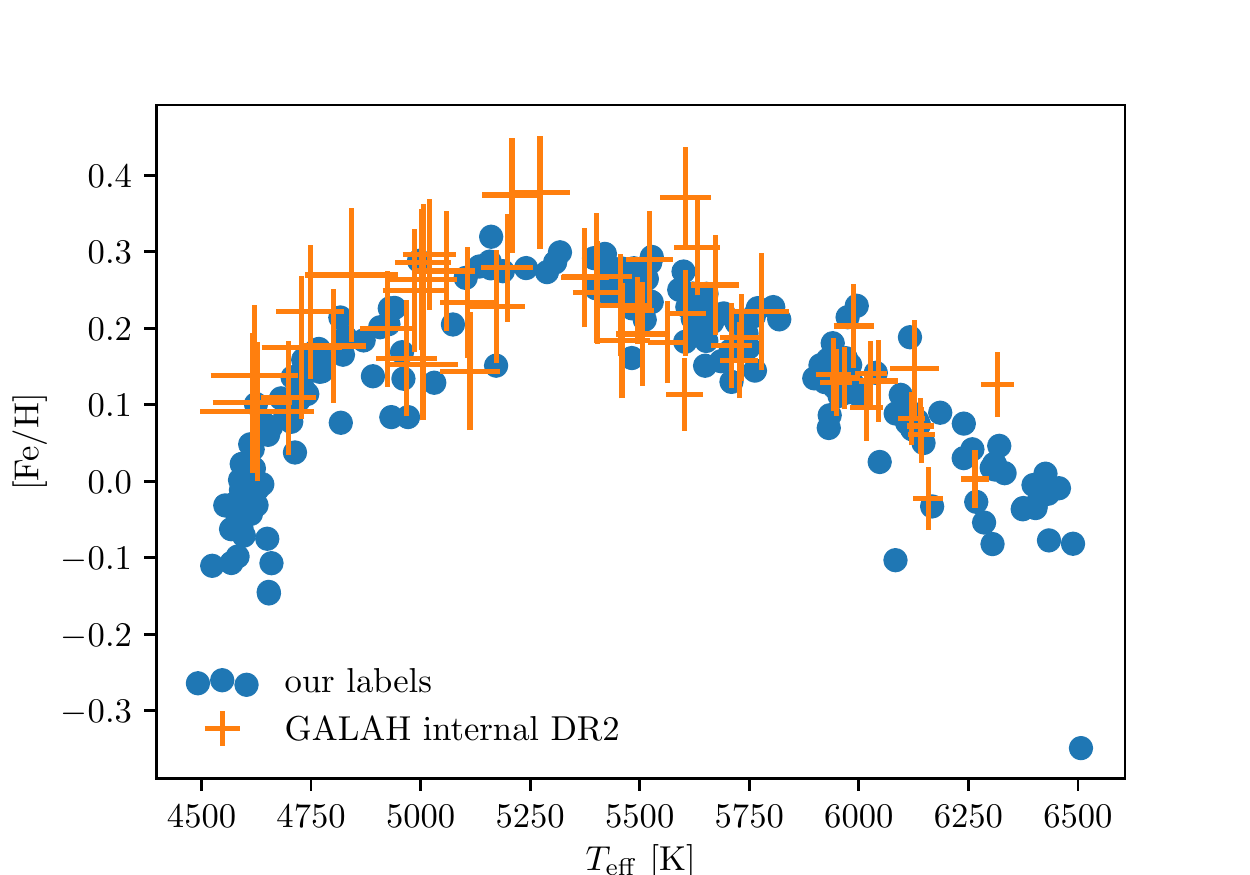}
        \caption{Inferred [Fe/H] vs effective temperature for \lamost{} and \galah{} dwarfs in Praesepe. 
        The \galah{} values come from internal DR2, which used the same analysis pipeline as the DR2 values in our training set.
        The systematic trend with $\teff$ is spurious, since all stars in Praesepe have the same abundances to below the precision achievable with low-resolution spectra.}
        \label{fig:praesepe}
    \end{figure}
    
    \item 
    Examination of open clusters in our catalog reveals that our inferred abundance ratios for dwarf stars are subject to strong systematics as a function of $\teff$.
    There are astrophysical explanations for weak abundance trends with $\teff$ and $\logg$, such as atomic diffusion \citep{dotter:17, gao:18, souto:19}, but not trends of this magnitude. 
    Similar systematics are present in the GALAH DR2 internal catalog (which employs the same analysis pipeline as the public release), as well as the official \lamost\ [Fe/H] values for dwarfs, suggesting that these trends are not introduced by our label transfer, but are present in \emph{ab-initio} stellar models and possibly inherited via our training set. 
    There are no obvious systematics in the red giant stars in our catalogue, save for [Ba/Fe], discussed below. 
    However, LAMOST does not contain enough red giants in known open clusters or wide binaries to determine the presence and magnitude of any systematics conclusively.
    
    Figure \ref{fig:praesepe} shows systematic trends in Praesepe in [Fe/H] as a function of $\teff$ by plotting our inferred values (for stars selected by \citealp{gaia_clusters:2018}) alongside GALAH internal DR2  values for the same open cluster, which is expected to be chemically homogeneous to a level well below our precision.
    \galah{} internal DR2 includes stars not part of the public DR2, but it employs the same analysis pipeline.
    Trends of the type exhibited in Figure \ref{fig:praesepe} are reduced but not eliminated in \galah{} DR3 (Buder et al. \emph{in prep}).
    Other abundances exhibit similar behavior. 
    This indicates that systematic error as a function of stellar parameters is a major contributor to our abundance error (see also Section \ref{sec:widebinaries}).
    To the extent that these trends are physical, the recommendation that the catalog user compare stellar abundances within a narrow range of $\teff$ remains.

    The systematic trends we see in dwarf abundances could be ``calibrated out'' using the nearly 3000 stars in LAMOST DR4 open clusters (with two or more targets), 
    \citep{cantatgaudin:18}, and 142 known wide binaries \citep{el-badry:19}.
    Instead of applying a \emph{post-hoc} correction, they could also be used to constrain the model at training time.
    Correcting for these systematics in the dwarf population is beyond the scope of our analysis.
    Despite this systematic effect, our abundances for dwarf stars are still useful for conducting analyses in restricted temperature ranges (see, for example, our examination of abundances of wide binaries in Section \ref{sec:widebinaries}).
    When examining the abundance trends across the disk, we exclude the dwarf stars, and focus on the $\approx$ 1 $\times$ 10$^6$ red giant stars in our catalogue.
    These giants span a vast spatial extent, and alone demonstrate the scientific potential of the distribution of stellar abundance data across the Galaxy.
\end{itemize}
 
Unless otherwise stated, in the sections below we employ stars in our catalog for which for which \texttt{chi2} is less than 7000. 
Other cuts were not found to have an effect on the results presented below. 

\subsection{Detailed abundances of wide binaries} \label{sec:widebinaries}
\begin{figure}
    \centering
    \includegraphics[width=0.45\textwidth]{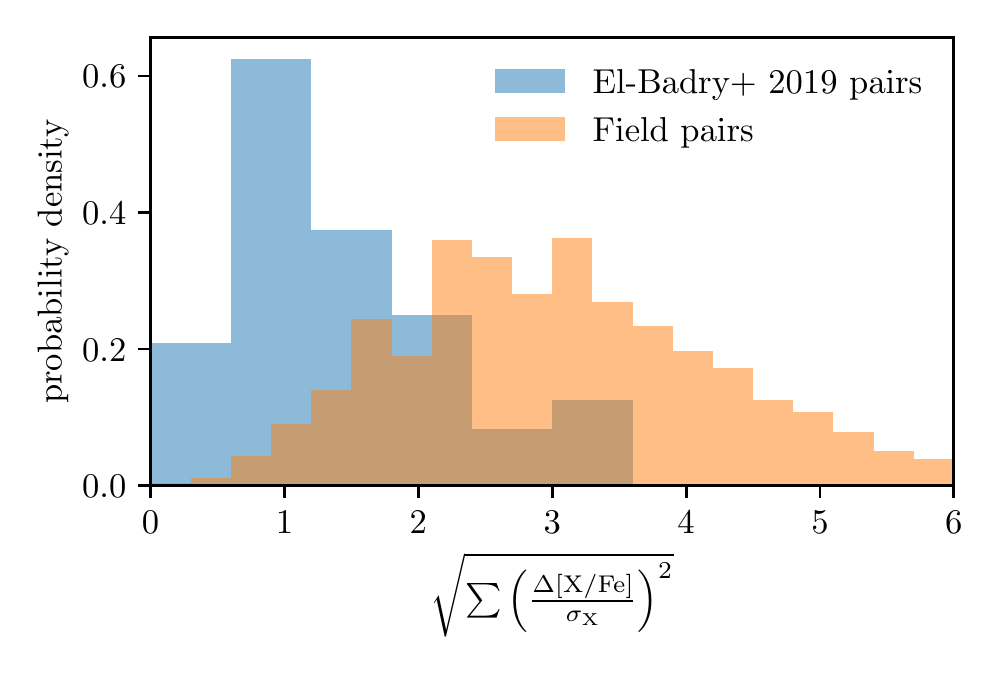}
    \caption{
    Error-weighted chemical distances for EB19 wide binaries and for random field pairs selected with the same cuts and chosen to have the same $\mathrm{\Delta[Fe/H]}$ distribution as the EB19 sample. 
    The wide binaries are more chemically similar than implied by their similarity in bulk metallicity alone.
    }
    \label{fig:pairs}
\end{figure}

\cite{el-badry:19} (hereafter EB19) used \gaia\ to identify wide binaries in the solar neighborhood and examined their properties as a function of [Fe/H].
We examined the detailed abundances of those present in \lamost.
For ease of analysis, we excluded pairs for which $\teff$ or $\logg$ were not available and those containing at least one giant (a total of 8 pairs).
Because of the strong systematic trends with $\teff$ that are present in our dwarf abundances, we constrain our analysis to wide binaries with $\Delta \teff < 250~\mathrm{K}$, for which both stars are LAMOST dwarfs.

To confirm the additional discriminating power of our inferred abundances, we examined the abundance similarity of wide binaries compared to a reference sample of non-binary pairs.
We constructed a set of random pairs of field stars, where each pair has the same metallicity as the binary pair.
The reference stars also conform to the quality cuts made in EB19 with $\teff < 250~\mathrm{[K]}$.
We used rejection sampling to ensure that they had as closely as possible the same $\Delta \mathrm{[Fe/H]}$ distribution as the EB19 sample. 
By comparing the abundance distribution of the random field pairs with the wide binaries, we can characterize the amount of information contained in our detailed abundances \emph{above and beyond} that contained by the bulk metallicity, $\feh$.
To capture the difference in chemistry between stars we use precision-scaled Euclidean distance,
\begin{equation}
\sqrt{\sum_i \left( \frac{\Delta \mathrm{[}X_i\mathrm{/Fe]}}{\sigma_i} \right)^2},
\end{equation}
where the $X_i$'s are the elements estimated, and the $\sigma_i$'s are their CV-assessed uncertainty.
Figure \ref{fig:pairs} shows the distribution of these chemical distances for both the wide binaries and the field pairs with the same $\Delta$[Fe/H] distribution. 
The difference between these distributions shows that detailed chemical abundances provide additional information about star's birth sites. 
Each abundance included pushes the chemical difference distribution of the binaries and random pairs further apart.  
The wide binaries peak at a smaller chemical distance than the reference pairs. Wide binaries peak at a distance of $~0.8$ and reference stars peak at a distance of $~2.5$.
This is consistent with findings that the majority of wide binaries are chemically identical to at least the $0.1$ dex level \citep{Andrews:2018, Andrews:2019, Hawkins:2019}.
We did not find that binaries with a larger separation are more chemically different, in contrast with the results of \citet{ramirez:19}.

If a cut in $\Delta \teff$ is not made, the systematic error in each abundance becomes similar in magnitude to the dispersion of chemistry in the solar neighborhood ($~0.1 - 0.5$ dex, depending on abundance).
Without this $\Delta \teff$ cut, and with this subsequent high systematic error, random field pairs and wide binaries appear to have very similar chemical difference distributions.
Even more stringent requirements for $\Delta \teff$ result in even more distinct chemical distance distributions for the wide binaries, but at the expense of the number of qualifying wide binaries.
In fact, the chemical differences we see in the wide binaries are much smaller than the error we get in cross-validation. This suggests that systematic $\teff$-dependent effects dominate our CV-assessed errors (see Figure \ref{fig:praesepe}).
If, in future work, we were able to reduce or eliminate this effect, perhaps by conditioning a model on chemically homogeneous open clusters, we could produce much higher-fidelity detailed abundances. Currently, scientific exploitation of our $\approx$ 3 million dwarf stars should employ narrow ranges of $\teff$.

Our tests of the chemical differences between wide binary stars indicate not only that the detailed chemistry provides evidence of a common birth site.
They also show that systematic effects are a large fraction of our error budget--a promising sign that we can do better with low-resolution spectra in the future.

We also similarly investigated the chemical differences for a sample of co-moving pairs in \citet{kamdar:19} compared to a reference set of field stars at the same $\feh$, and found that they were also chemically more similar than an equivalent set of field pairs, although less so than the wide binaries.
\citealp{simpson:19} used \galah\ abundances to determine whether 15 co-moving pairs found in \gaia\ were co-natal; the same approach could be used here.


\subsection{Mapping chemistry in the Milky Way} \label{sec:maps}
\begin{figure*}
    \centering
    \includegraphics[width=\textwidth]{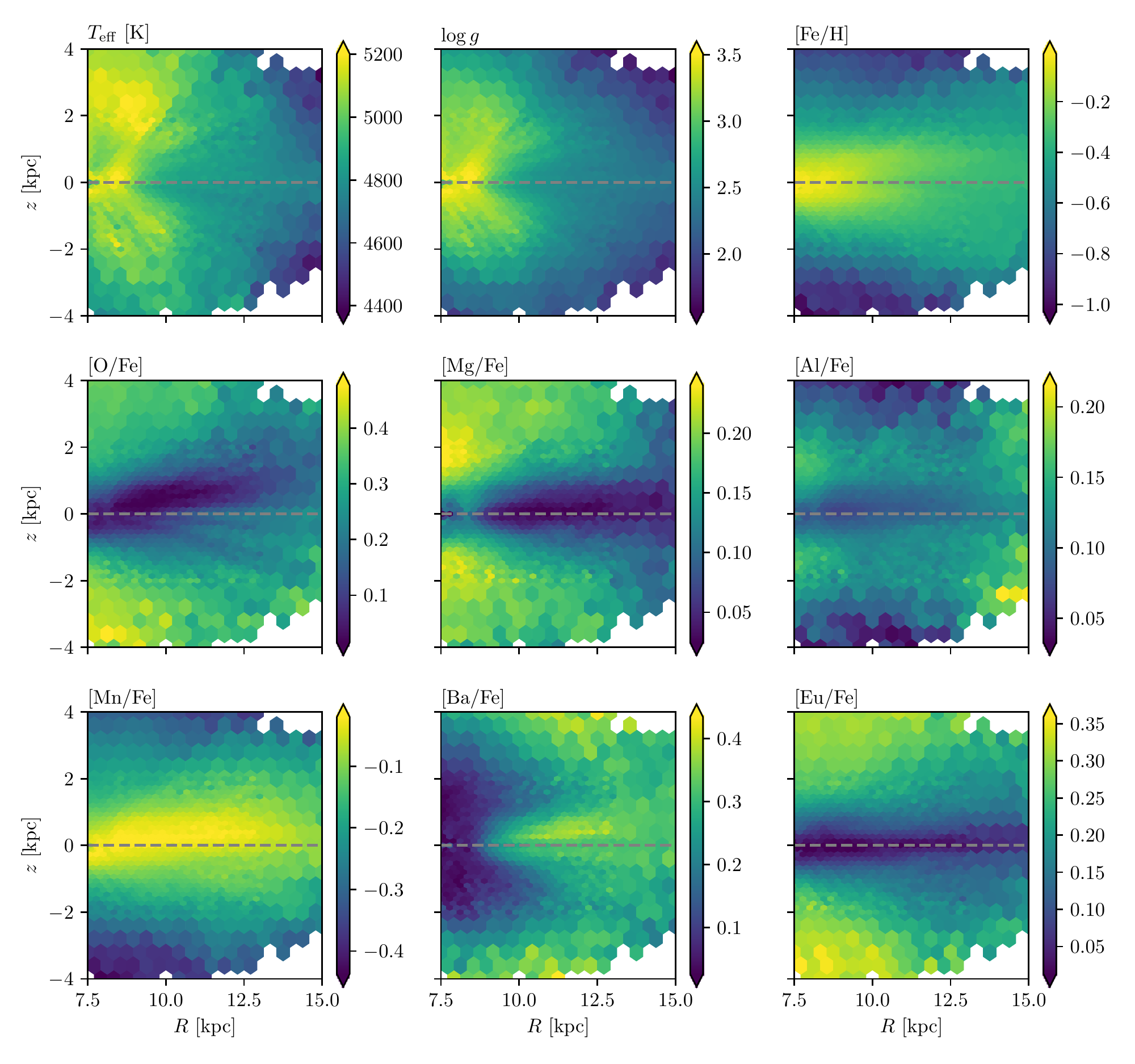}
    \caption{
    The $(R,z)$ plane colored by mean label value for nine labels of our 800,000 giant stars, summed over all $\phi$.
    The flaring of the disk can be seen in [Eu/Fe], [Mg/Fe], and [Mn/Fe].
    }
    \label{fig:profile}
\end{figure*}
We have one of the largest homogeneous samples of stellar abundances.
This sample is ideal for mapping the abundance distribution of the Milky Way across a large spatial extent.
First, we map the disk across the meridional plane, (R,z), to characterize the spatial abundance trends in that plane.
Similar maps can be created with a different set of abundances using \apogee\, but that data set is most concentrated to the disk and the inner Galaxy, while the \lamost\ giants more extensively span the halo and outer disk. 
\apogee\ data clearly reveal the flaring in intermediate-age populations in the $(R,z)$ plane (e.g. \citealp{ness:16}).
This is presumably a consequence of radial migration \citep[e.g.][]{roskar:08}, whereby stars increase in scale height as they move outward in the disk \citep{minchev:12}.
Due to the correlations between abundances and ages \citep{bedell:18, feuillet:18, feuillet:19, ness:19}, we might expect to also see such flaring in mean-abundance maps, although this is potentially confounded by the metallicity dependence of the age-abundance relationships \citep{ness:19}. 
Detailed analyses of the chemodynamical distribution across $(R,z)$ that seek to make any quantitative claims require a careful consideration of the \lamost\ selection function.
Characterisation of the flaring profile of the disk also require stellar ages, as noted by \citet{minchev:14, minchev:18}.
Here, we aim to show the potential of these data for more in-depth analysis, that  accounts for the selection function.

Figure \ref{fig:profile} shows the $(R,z)$ plane colored by mean label value for nearly 800,000 giant stars, for abundance ratios of Fe, O, Eu, Mg, Al, Mn, and Ba as well as $\teff$ and $\logg$.
These maps span $-4~\kpc < z < 4~\kpc$ and $7.5 ~\kpc < R < 15~\kpc$.
The disk is clearly distinct from the halo.
At $R = 8~\kpc$, for example,  the halo transition appears as a smooth mean abundance change centered on $|z| \approx 2~\kpc$.
Flaring is seen in the individual elements, particularly for O, Mg, Eu and Ba.
All of these elements show different flaring, of varying strength and profile.
All abundances increase or decrease monotonically with $|z|$ at fixed $R$, except for [Al/Fe], which increases  with $|z|$ until $|z| \approx 2~\kpc$, beyond which it decreases with $|z|$.

The apparent barium-depleted ``cone'' centered on the sun is caused by systematic trends in [Ba/Fe] as a function of $\teff$  and $\logg$, in combination with \lamost's selection function. 
If we plot only the red clump stars as identified by \citet{ting:2018_rcs} (roughly $2\times10^5$ stars), which exhibit a narrow range of stellar parameters and which have very precise photometric distances, this feature disappears. The shape and morphology of flaring in the elements is preserved when examining the red clump stars only.
Finally, we note that both [Ba/Fe] and [O/Fe] appear to be asymmetrically distributed about the Galaxy's midplane.
This asymmetry in the mean abundance value around the midplane persists in maps of the $(R,z)$ plane made with only red clump stars, suggesting that they are not related to $\teff$-dependent systematics.
As seen in Figure \ref{fig:profile}, this feature doesn't correlate clearly with $\teff$ or $\logg$; nor does it trace extinction as traced by dust maps, or mean signal to noise of the stars.
We do not rule out the possibility that the midplane asymmetry seen in these elements is caused by selection effects, particularly in light of the fact that these asymmetrical features are stretched along lines of sight.

\begin{figure*}
    \centering
    \includegraphics[width=\textwidth]{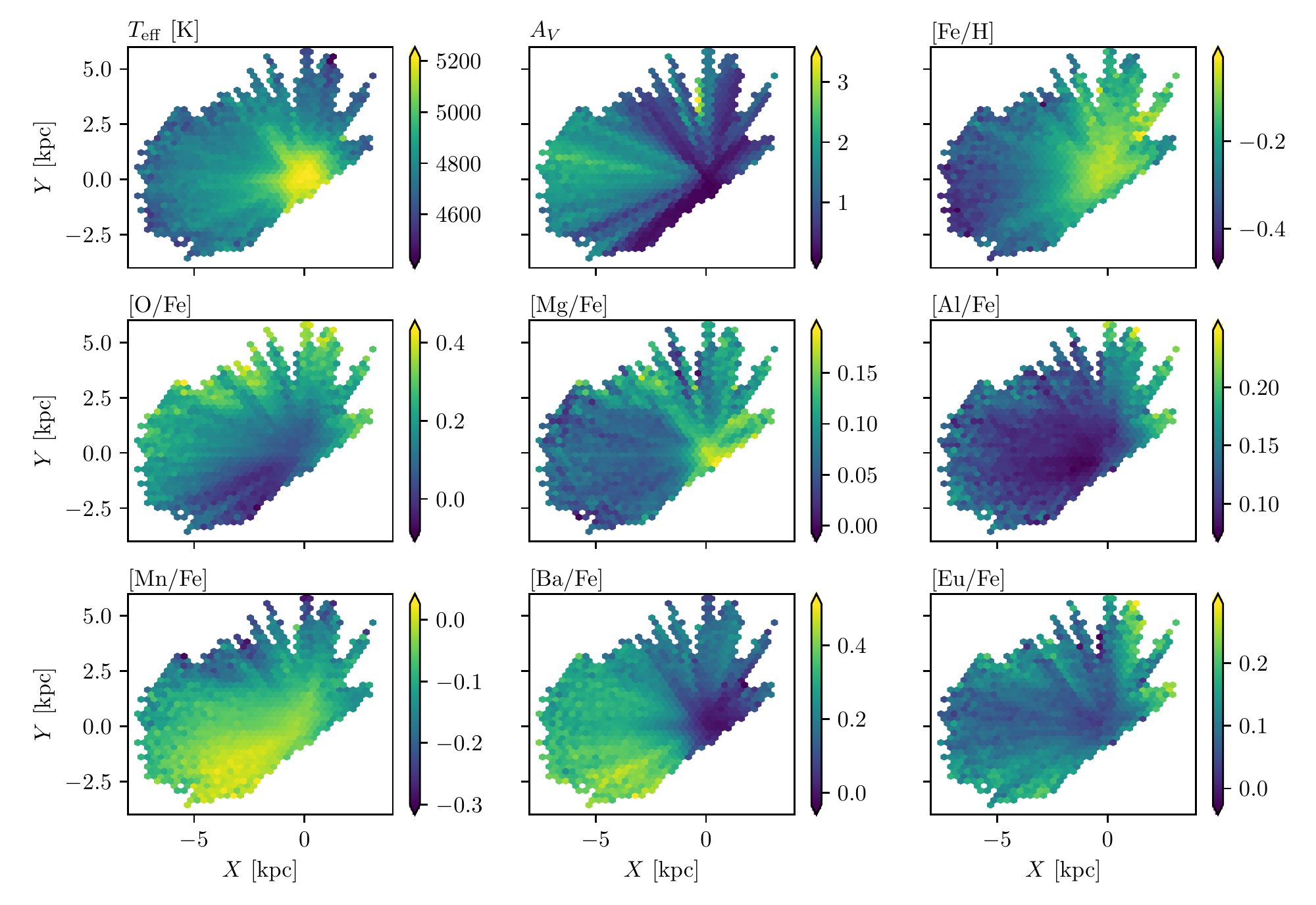}
    \caption{Mean abundances of \lamost\ giants in the thin disk ($J_z < 30~\mathrm{kpc~km~s^{-1}}$), mapped in the $(X,Y)$ plane.
    $\teff$ and $A_V$ are also mapped for comparison.
    }
\label{fig:azimuthal}
\end{figure*}

Figure \ref{fig:azimuthal} shows mean-abundance maps in the $(X,Y)$ plane for kinematic thin disk stars ($J_z < 30~\mathrm{kpc~km~s^{-1}}$), as well as mean $\teff$ and $A_V$ maps, for comparison.
We highlight the apparent azimuthal structure in [Mn/Fe] and [O/Fe], which isn't easily explained by spurious correlation with $\teff$, $A_V$, S/N, $z$, or $J_z$.
These are two abundances for which our inferred values have lower relative uncertainty (Figure \ref{fig:reluncert}).
Again, we note that a complete treatment of these data would involve explicit modelling of the \lamost\ selection function.
Note that the sensitivity of [Ba/Fe] to $\teff$ is clearly visible.
Like the ``cone'' in the $(R, z)$ plane, correlation of [Ba/Fe] with heliocentric distance disappears in maps including only red clump stars.
The patterns in these maps along lines of sight are likely of an observational origin, but are not easily explained by a single confounding factor.  
They do not trace mean height, extinction, metallicity, $\teff$, or $\logg$.

Azimuthal trends in abundance are known to exist in Galactic gas, and are often attributed to spiral structure (e.g. \citealp{Wenger:19}).
Variations in the height of the  midplane in combination with \lamost's selection function could give rise to azimuthal abundance gradients, but it is not clear why this would be manifest in some abundances and not others unless due to  abundance-age correlations, which does not appear to be the case here.
Additionally, there is no correlation between the strength of vertical (Figure \ref{fig:profile}) and azimuthal (Figure \ref{fig:azimuthal}) gradients, which would be expected if the azimuthal trends were due to midplane variations.

\begin{figure*}
    \centering
    \includegraphics[width=\textwidth]{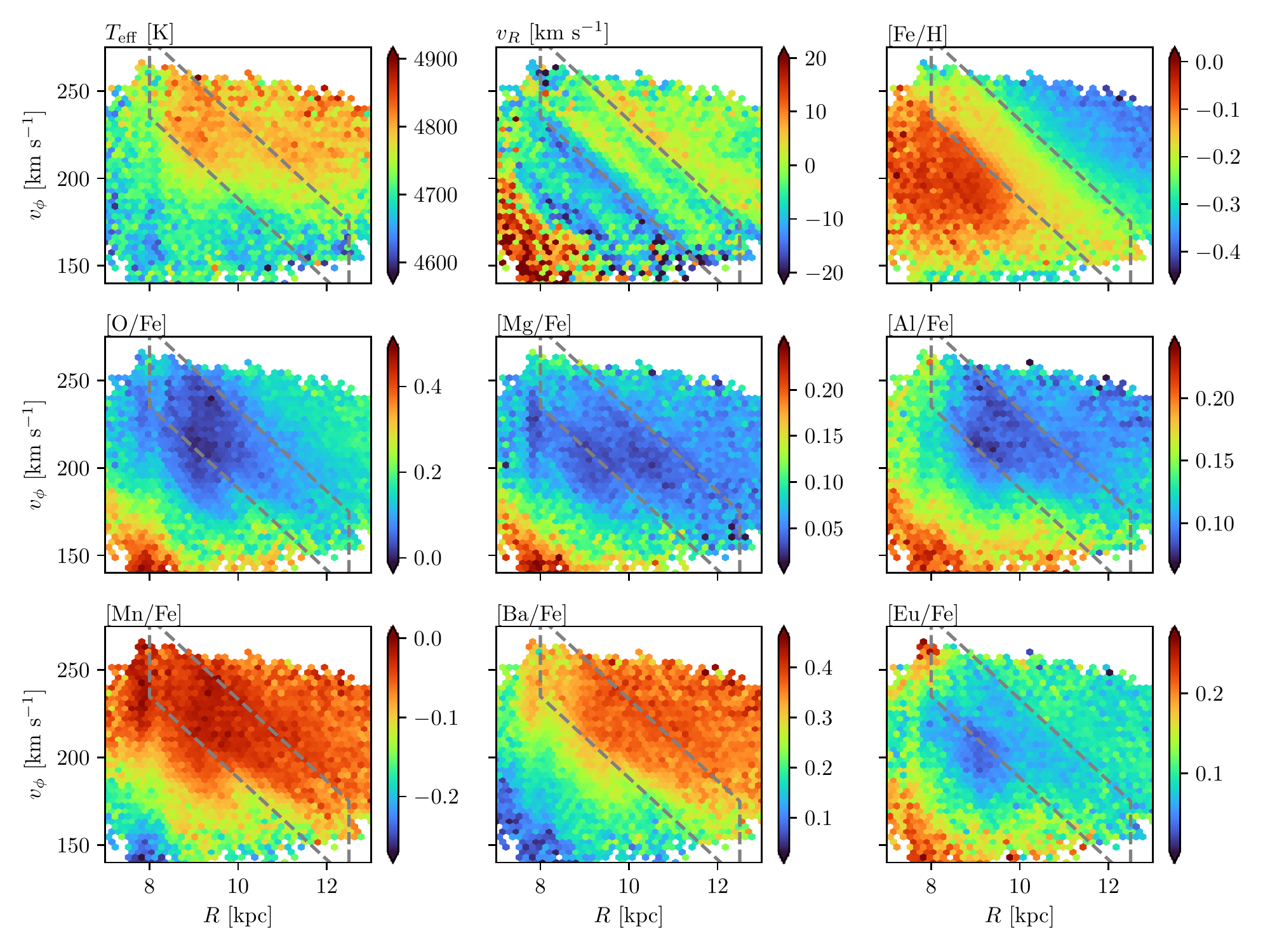}
    \caption{Mean-abundance maps of thin disk ($J_z < 30~\kpckms$) \lamost\ giants in the $(R, v_\phi)$ plane.
    The dashed box outlines the longest ridge of increased stellar number density.
    We plot only stars with $\logg < 3$ to minimize contamination from dwarfs, whose abundances are generally on a different scale and have $\teff$-dependent systematic trends, but a vertical feature is still visible at the solar radius, $R = 8~\kpc$, presumably induced by the selection function.
    }
    \label{fig:ridge}
\end{figure*}

Figure \ref{fig:ridge} shows mean-abundance maps of \lamost\ giants for the thin disk ($J_z < 30~\kpckms$) in the $(R, v_\phi)$ plane.
By using \lamost\ radial velocities, we are able to probe further out into the disk than with the Gaia DR2 RVS sample.
In this plane a particularly prominent feature are the `ridges’ first reported by \citet{Kawata:2018}.
A number of interpretations have been given for the origin of these ridges, including perturbations introduced by spirals, the bar, an external perturber or a combination of these (e.g. \citealp{Antoja:2018, Khanna:2019, Bland-Hawthorn:2019, Fragkoudi:2019, Fragkoudi_chemo:2019, Laporte:2019}).
Of particular interest is the longest ridge (outlined by a dashed line).
Wheeler et al. 2019 (\emph{in prep}) will discuss its dynamical origin.
Of our abundances, the ridge is most visible in the [O/Fe] and [Mn/Fe].
These are two elements that display the clearest azimuthal abundance gradients, and which our inferred abundances have the lowest relative uncertainty (see Figure \ref{fig:reluncert}).

\subsection{The disk-halo transition seen in chemistry}

\begin{figure*}
    \includegraphics[width=\textwidth]{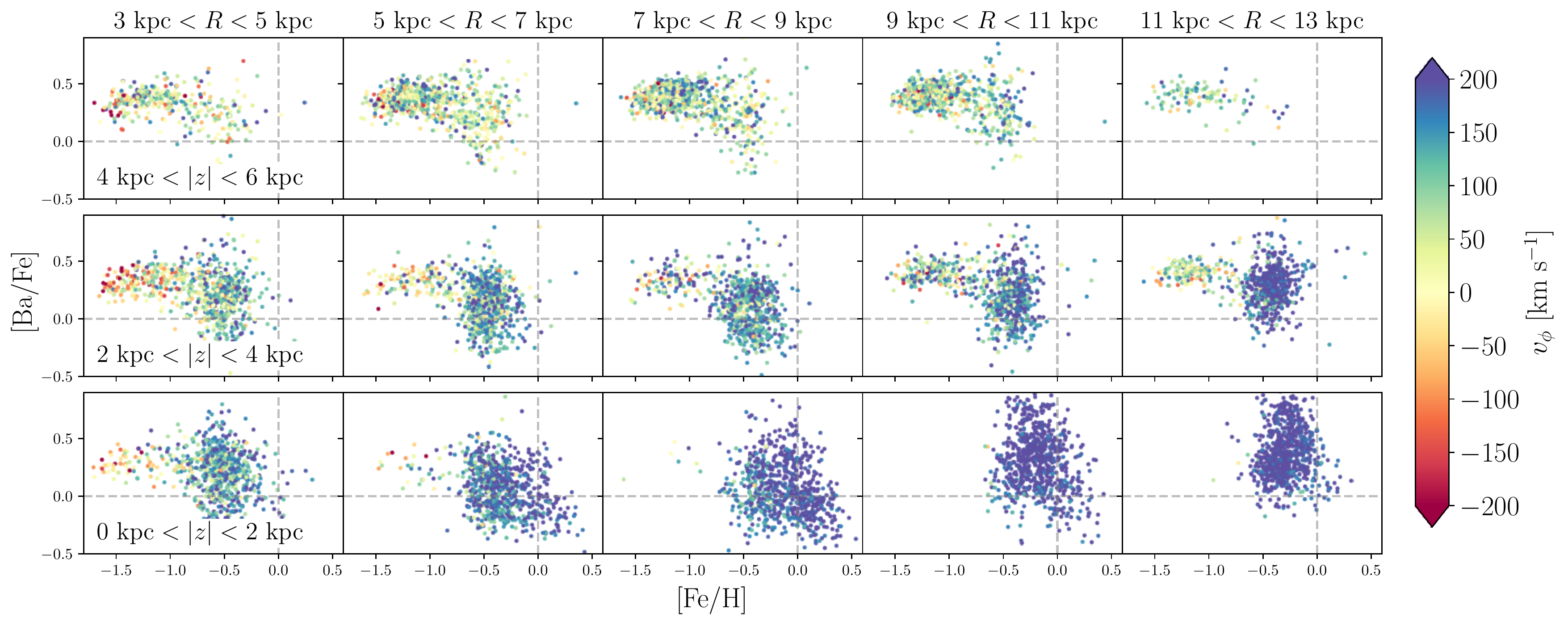}
    \caption{
    The ([Ba/Fe]-[Fe/H]) plane, colored by azimuthal velocity, $v_\phi$, and plotted in spatial bins in the Galaxy, with (up to) 700 randomly selected stars plotted in each bin.
    Because of the $\teff$-dependent systematics in our inferred [Ba/Fe] values, we have only plotted stars with $4800~\mathrm{K} < \teff < 5000~\mathrm{K}$.
    With increasing $|z|$, halo stars (as distinguished by their lower $\phi$-velocities) become more dominant.
    }
\label{fig:hayden1}
\end{figure*}

\begin{figure*}
    \includegraphics[width=\textwidth]{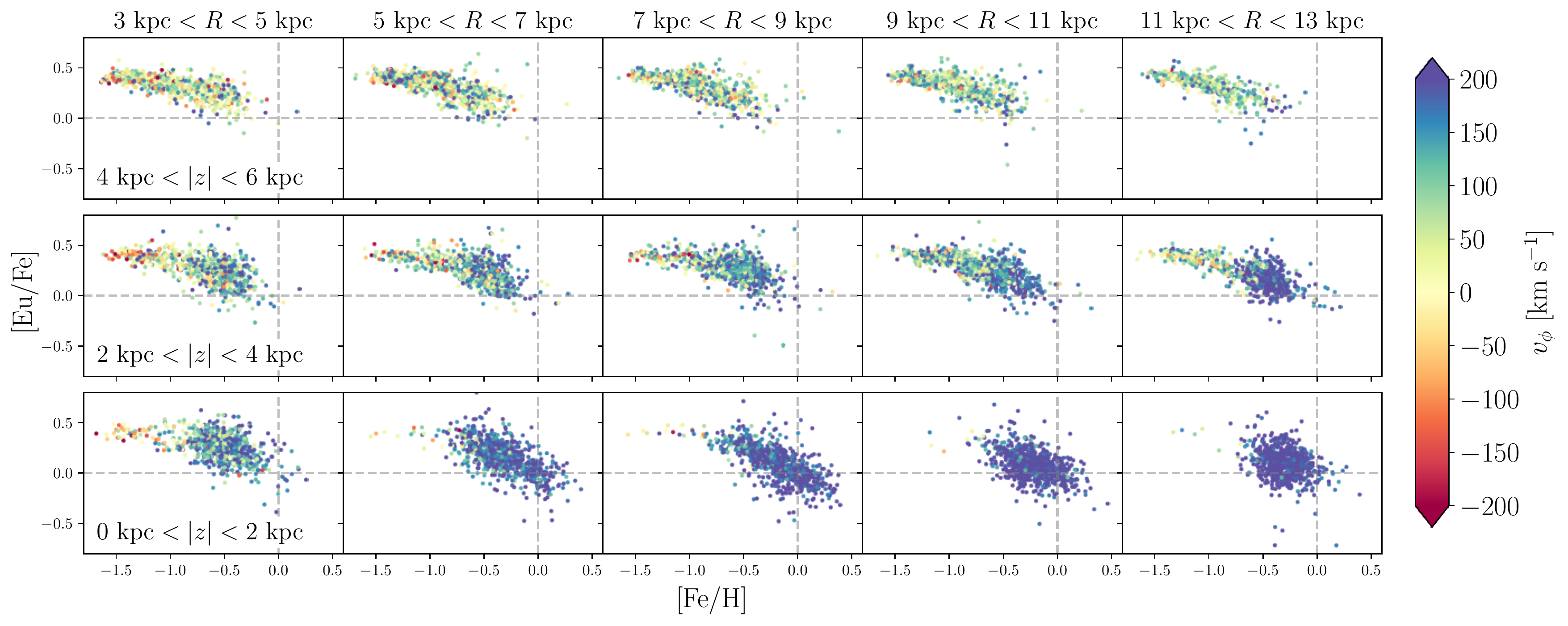}
    \caption{
    Same as figure \ref{fig:hayden1}, with [Eu/Fe] in place of [Ba/Fe] and with stars chosen without restriction on $\teff$.
    }
\label{fig:hayden2}
\end{figure*}

The distributions of the neutron capture elements, [Ba/Fe] and [Eu/Fe], vs [Fe/H], colored by $v_\phi$, are plotted in spatial bins in Figures \ref{fig:hayden1} and \ref{fig:hayden2}.
Because of the $\teff$-dependent systematics in [Ba/Fe], we have only plotted stars with $4800~\mathrm{K} < \teff < 5000~\mathrm{K}$ in Figure \ref{fig:hayden1}.
Using other temperature ranges doesn't qualitatively change the plot, but not restricting $\teff$ yields higher dispersion in [Ba/Fe]. The azimuthal velocity,
$v_\phi$, allows us to clearly distinguish between the disk and halo populations at the scale heights (and vertical actions, $J_z$) where both are present (primarily the center row in Figures \ref{fig:hayden1} and \ref{fig:hayden2}), illuminating the chemical differences between them. 
Disk stars are prograde across $R$, and concentrated to low $z$, with most disk stars having $v_\phi \gtrsim 100~\mathrm{km~s^{-1}}$.
There are fewer metal-poor disk stars as $R$ increases, with a narrower more metal rich distribution at larger $R$, seen across the smallest $z$ range.
Halo stars are seen at larger $z$ and are characterized by their more isotropic, eccentric orbits.
At $11~\kpc < R < 13~\kpc$, most halo stars have $v_\phi \lesssim 200~\mathrm{km~s^{-1}}$ with a distribution of $v_\phi = 80 \pm 70~\mathrm{km~s^{-1}}$.
The halo stars also appear to have increasingly negative velocities in the inner Galaxy. At our intermediate height from the plane, $2~\kpc < z < 4~\kpc$, the metal-poor stars ($\feh < -1.0$) are predominantly retrograde at our smallest $R$ range, $3~\kpc < R < 5~\kpc$. 
Cutoffs of $J_R \approx 100 \mathrm{km~s^{-1}~kpc}$ and $J_\phi \approx 1500~\mathrm{km~s^{-1}~kpc}$ also clearly distinguish between disk and halo stars at large $R$. 
\reviewer{These populations are not as dramatically distinguished at small $R$ because distance errors propagate to larger uncertainties in $v_\phi$ (see Appendix \ref{sec:kinematicerrors}).}

The distribution of (kinematic) halo stars in the ([Ba/Fe],[Fe/H]) plane has a transition at $\mathrm{[Fe/H]} \approx -1$ (most clearly seen in the middle row of Figure \ref{fig:hayden1}, $2 ~\kpc <  |z| < 4~\kpc$).
This metallicity corresponds to the transition between the disk and halo, as well as the approximate boundary between the accreted and (at least one component of the) \insitu\ halo (\citealp{bonaca:17}, called ``the splash'' in \citealp{belokurov:19} and the ``heated thick disk'' by \citealp{dimatteo:18}).
At least for barium, the abundance planes at $4~\kpc< |z| < 6~\kpc$, suggest an overlap in the chemical plane of different sequences, perhaps associated with the accreted and \insitu\ halo.
Both [Eu/Fe] and [Ba/Fe] have larger dispersion at high [Fe/H], but the sequence of europium abundances varies less across $z$.

\section{Discussion \& Conclusions}
We have trained a data-driven model (\thecannon)\ to estimate detailed abundances from low-resolution LAMOST spectra, delivering up to 7 abundances for $3.9 \times 10^6$ stars to a precision of $0.05 - 0.23$ dex.
$2.9 \times 10^6$ of these are dwarf stars, for which we infer labels with $0.05 - 0.23$ dex precision.
$8.8 \times 10^5$ are red giants for which we infer abundances with $0.07 - 0.22$ dex precision.
Our best-fit model spectra are easily reproducible using our catalog, implementation of \thecannon,
and model coefficients\footnote{See \href{https://doi.org/10.7910/DVN/5VWKMC}{doi.org/10.7910/DVN/5VWKMC}}, which are available online.
We used the red giants to examine the spatial distribution of abundances in the disk and halo and the dwarf stars to investigate the chemical similarity of wide binaries.

Our analysis of the chemical similarity of dwarf stars in wide binaries compared to field stars showed these stars are from a common birth site and enabled us to quantify the additional resolving or discriminating power in the vector of our derived abundances beyond an overall metallicity. 

Using the red giants, we first mapped the profile of the disk in the $(R,z)$ plane in the elements O, Eu, Mg, Al, Mn, and Ba. These maps show the flaring of the disk and the distinction in abundances between the halo population at high latitudes and the disk. 
Second, we examined face-on projections across the disk in this set of abundances.
These projections hint at some non-axisymmetric patterns in the abundances.
Indeed, the \gaia\ mission has revealed numerous dynamical deviations from axisymmetry in the disk and perturbations in the solar neighbourhood \citep{Antoja:2018, Sellwood:2019, Trick:2019}.
Third, we constructed mean-abundance maps in the $(R, v_\phi)$ plane and discuss the chemical signature of the high density ridges in this plane.
Finally, we investigated the abundance planes of [Ba/Fe]-[Fe/H] and [Eu/Fe]-[Fe/H] across $(R,z)$, making similar maps to \citet{hayden:15}, but with the neutron capture abundances.
These maps showed the disk and halo trends across $\feh$ at all $(R,z)$. 
These different trends might be used to separate any in-situ halo from heated disk stars, from an accreted halo.
As the set of abundances we deliver give higher discriminating power to identify chemically similar stars compared to $\feh$ alone, we expect the multiple families of abundances will be useful for studies of the plethora of chemodynamical substructure in the Milky Way halo (see e.g. \citealp{belokurov:19, helmi:18, dimatteo:18, myeong:19, Antoja:2018}).

We derive abundances with diverse nucleosynthetic channels and are demonstrably uncovering some of the breadth of chemical information in the Milky Way.
However, a number of caveats are discussed in Section \ref{sec:cat} and we further detail some of these here.

In contrast to \citet{ho:17a, ho:17b}, we find that our cross validation results do not vary strongly with S/N.
This indicates that our precision is limited by that of reference labels themselves, and, if improved, we would obtain higher precision results for our test objects that scale as expected with SNR.

Because our model is in some cases not inferring abundances from the corresponding lines themselves, it may not be robust to stars with properties or enrichment histories not represented in the training set.
This means that, while stars that are highly enriched or depleted in an element may not have their abundances accurately inferred, the best fit model should have large residuals in regions of its known lines.
\citet{casey:19} used this effect to identify stars that are unusually rich in lithium, but this approach could be extended to all elements with strong lines in the \lamost\ wavelength range. 
In particular, it is challenging to measure $r$-process abundances from $r$-process absorption features even from extremely high quality high resolution spectra, and GALAH uses only 2 relatively unblended absorption regions for their [Eu/Fe] measurement (Table \ref{tab:windows}).
It raises questions about stellar spectra that we are inferring [Eu/Fe] (albeit noisily) from relatively low-$S/N$, low-resolution spectra, as confirmed by cross validation, and that the [Eu/Fe] distribution across [Fe/H] mirrors that of boutique studies (e.g. \citealp{Bensby2005}).
The physical origin of the significant correlations between absorption features of nominally different nucleosynthetic families \citep{Feeney:2019} is not clear, but is presumably caused by a combination of the inherent correlation induced by element-production mechanisms, shared chemical enrichment history, and stellar physics. 
In other words, the chemical manifold on which the majority of stars lie is not well-known.
Of significant interest is most likely those stars where we can not well match the spectra with our data-driven model, which is by far the minority of stars in \lamost.

The large number of \reviewer{low- and medium-}resolution spectra available now (RAVE and SEGUE, in addition to \lamost) and coming in the near future (e.g. WEAVE, \citealp{dalton:12}, MOONs, \citealp{cirasuolo:14}, and 4MOST, \citealp{dejong:19} \reviewer{in their lower-resolution modes}; DESI, \citealp{desi:16}, Sloan V \citealp{Kollmeier:2017}, \reviewer{and \emph{Gaia}, \citealp{gaia:16}}) makes honing our ability to learn from these data a fruitful endeavor.
We also discussed future improvements to our methodology, especially the possibility of using open clusters to reduce the effect of systematic trends with stellar parameters in inferred abundances.
Other promising methodological directions include using more robust inference for model parameters and labels, perhaps allowing more rigorous error estimation, and allowing missing labels in the training set, which would enable us to used training data from multiple surveys.

\software{
Optim.jl \citep{mogensen2018optim}, Matplotlib \citep{Hunter:2007}, galpy \citep{bovy:15}
}

\section*{Acknowledgments}
The authors thank Kathryn Johnston, James Applegate, David Helfand, Tomer Yavetz, Klemen Čotar, Francesca Fragkoudi, and Matthew Abruzzo for useful discussion and notes. We thank Matthias Steinmetz for his constructive comments that have improved the paper. 

AJW is supported by the National Science Foundation Graduate Research Fellowship under Grant No. 1644869.

MKN is in part supported by a Sloan Research Fellowship. We thank and acknowledge the Kavli Institute for Theoretical Physics at the University of California, Santa Barbara, and in particular the Gaia19 workshop. This work was also in part inspired during the Aspen Center for Physics `Dynamics of the Milky Way System in the Era of Gaia' workshop, which is supported by National Science Foundation grant PHY-1607611.

SB acknowledges funds from the Alexander von Humboldt Foundation in the framework of the Sofja Kovalevskaja Award endowed by the Federal Ministry of Education and Research. This research has been supported by the Australian Research Council (grants DP150100250 and DP160103747). Parts of this research were supported by the Australian Research Council (ARC) Centre of Excellence for All Sky Astrophysics in 3 Dimensions (ASTRO 3D), through project number CE170100013.

JBH is funded by an ARC Laureate Fellowship.

DBZ and JDS acknowledge the support of the Australian Research Council through Discovery Project grant DP180101791

TZ acknowledges the financial support of the Slovenian Research Agency (core funding P1-0188)

This research was supported in part by the National Science Foundation under Grant No. NSF PHY-1748958.

Guoshoujing Telescope (the Large Sky Area Multi-Object Fiber Spectroscopic Telescope LAMOST) is a National Major Scientific Project built by the Chinese Academy of Sciences. Funding for the project has been provided by the National Development and Reform Commission. LAMOST is operated and managed by the National Astronomical Observatories, Chinese Academy of Sciences. 

We acknowledge computing resources from Columbia University's Shared Research Computing Facility project, which is supported by NIH Research Facility Improvement Grant 1G20RR030893-01, and associated funds from the New York State Empire State Development, Division of Science Technology and Innovation (NYSTAR) Contract C090171, both awarded April 15, 2010.

This work has made use of data from the European Space Agency (ESA) mission {\it Gaia} (\url{https://www.cosmos.esa.int/gaia}), processed by the {\it Gaia}
Data Processing and Analysis Consortium (DPAC, \url{https://www.cosmos.esa.int/web/gaia/dpac/consortium}). Funding for the DPAC has been provided by national institutions, in particular the institutions participating in the {\it Gaia} Multilateral Agreement.

This publication makes use of data products from the Wide-field Infrared Survey Explorer, which is a joint project of the University of California, Los Angeles, and the Jet Propulsion Laboratory/California Institute of Technology, and NEOWISE, which is a project of the Jet Propulsion Laboratory/California Institute of Technology. WISE and NEOWISE are funded by the National Aeronautics and Space Administration.

\bibliography{references}

\appendix

\section{Linear Coefficients} \label{app:coeffs}
\begin{figure}
    \centering
    \includegraphics[width=0.6\textwidth]{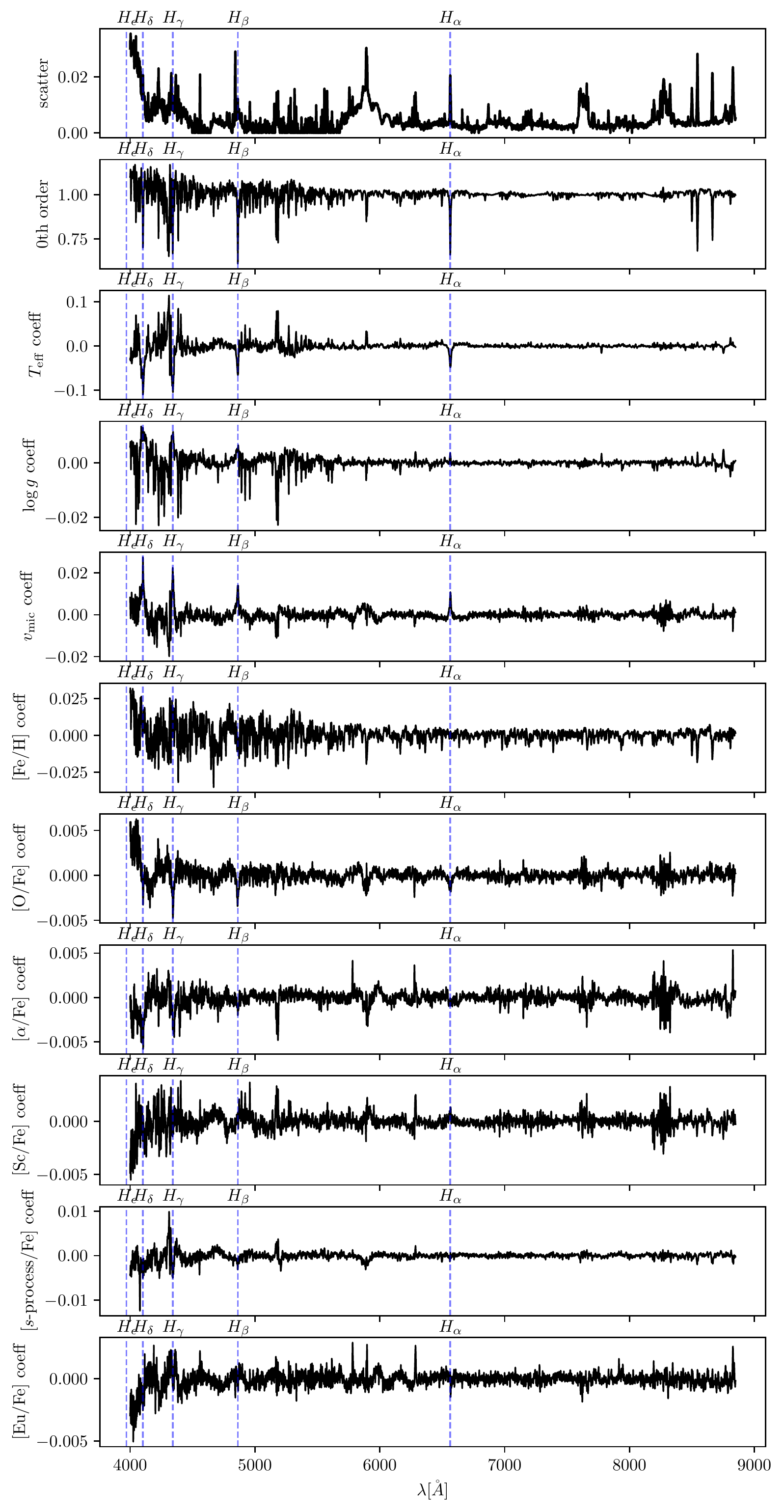}
    \caption{Linear coefficients for the dwarf model, omitting high-uncertainty pixels with $\lambda < 4000~\mathrm{\AA}$.}
   \label{fig:coeffs1}
\end{figure}

\begin{figure}
    \centering
    \includegraphics[width=0.6\textwidth]{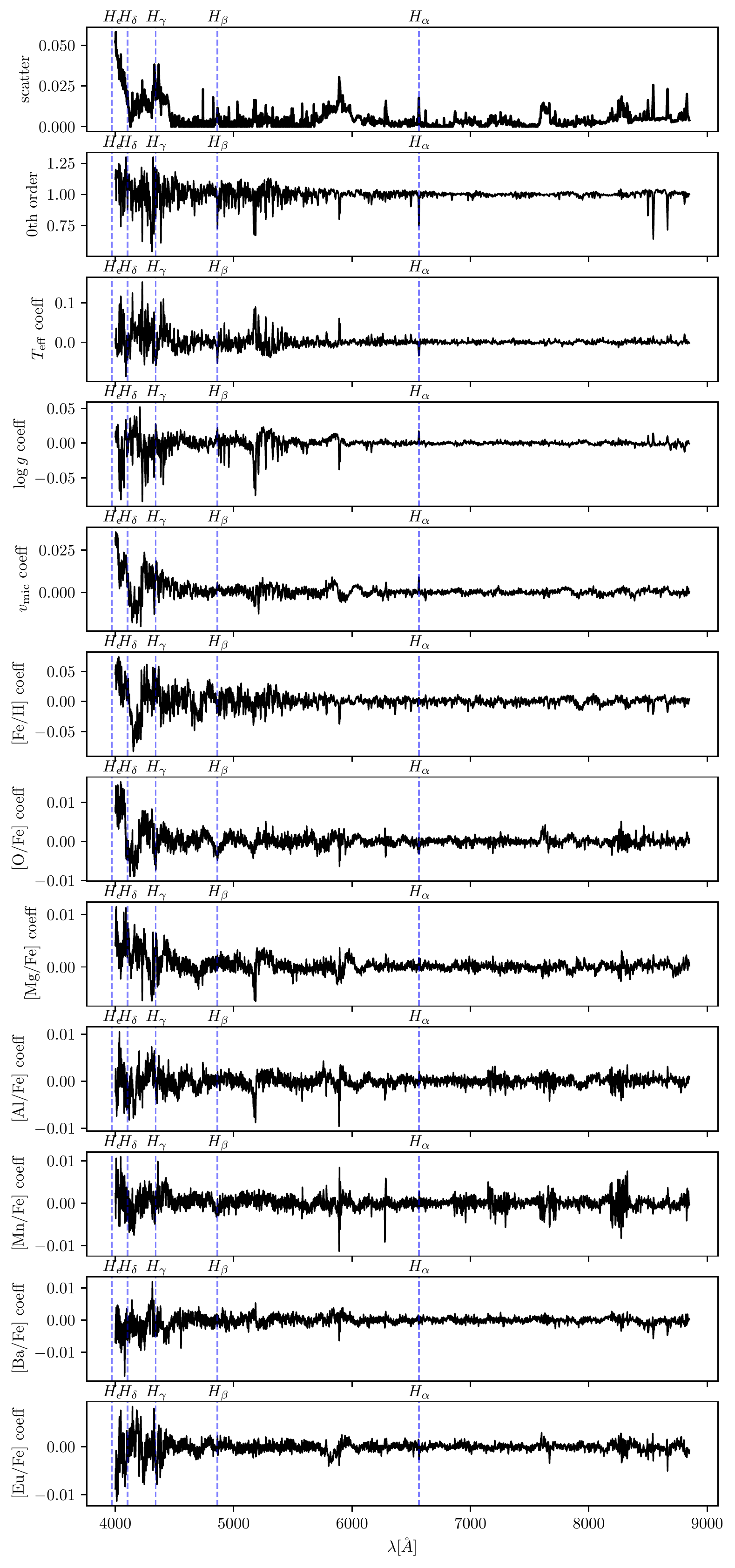}
    \caption{Same as Figure \ref{fig:coeffs1}, but for our giant model.}
    \label{fig:coeffs2}
\end{figure}

Figures \ref{fig:coeffs1} and \ref{fig:coeffs2} show the linear coefficients of our models as a function of wavelength.

\section{Radial velocity precision}
\begin{figure}
    \centering
    \includegraphics[width=0.6\textwidth]{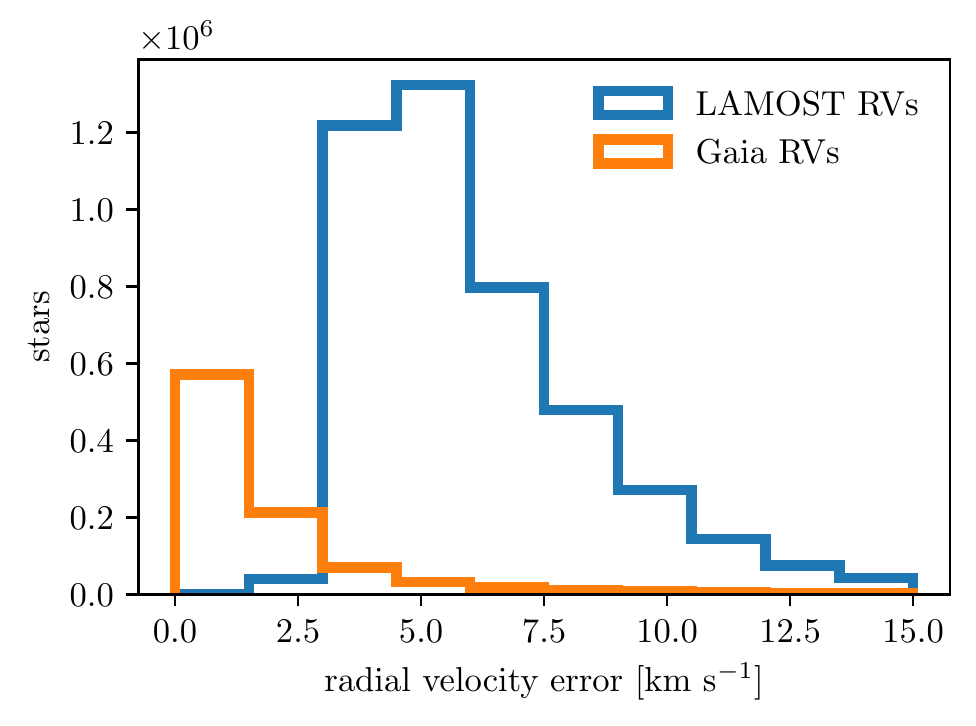}
    \caption{Reported RV precision of stars in our catalog from \gaia, and \lamost. While the \lamost\ precision is worse, \gaia\ only measued RVs for approximately one fifth of our catalog.}
    \label{fig:RVprec}
\end{figure}

Figure \ref{fig:RVprec} shows reported RV error for stars in our catalog, as measured by \gaia, and \lamost.
{}

\section{\galah\ windows} \label{sec:windows}
Table \ref{tab:windows} lists the \galah\ line windows for each estimate element.  We used these windows to calculate per-element $\chi^2$ values for \galah\ spectra to eliminate spurious measurements from our training set.

\begin{deluxetable}{lp{15cm}}
\tablecaption{ 
\galah\ line windows for each element
}
\label{tab:windows}
 \tablehead{
 \colhead{element} & \colhead{windows [\AA]} 
 }
\startdata 
Al & (6695.78, 6696.17), (6698.41, 6698.92), (7834.95, 7835.47), (7835.84, 7836.43) \\
Ba & (5853.53, 5853.86), (6496.68, 6497.19) \\
Ca & (5857.22, 5857.60), (5867.28, 5867.72), (6493.48, 6493.99), (6499.37, 6499.94), (6508.52, 6509.03) \\
Co & (6632.23, 6632.81), (7712.41, 7713.07), (7837.76, 7838.50) \\
Cr & (4775.03, 4775.21), (4789.20, 4789.47), (4800.83, 4801.20), (4847.98, 4848.31), (5702.12, 5702.50), (5719.50, 5719.99), (5787.64, 5788.14), (5844.40, 5844.79), (6629.80, 6630.25) \\
Cu & (5781.92, 5782.42) \\
Eu & (5818.61, 5818.99), (6644.97, 6645.29) \\
K & (7698.57, 7699.31) \\
La & (4716.29, 4716.61), (4748.62, 4748.85), (4803.92, 4804.24), (5805.52, 5805.96) \\
Li & (6707.37, 6708.26) \\
Mg & (4729.90, 4730.22), (5710.86, 5711.30) \\
Mn & (4739.01, 4739.29), (4761.37, 4761.64), (4765.69, 4766.06), (4783.17, 4783.58) \\
Na & (4751.71, 4751.94), (5682.54, 5682.92), (5687.93, 5688.37) \\
Ni & (5748.21, 5748.59), (5846.82, 5847.21), (6482.60, 6483.05), (6532.58, 6533.10), (6586.02, 6586.47), (6643.37, 6643.94), (7713.89, 7714.48), (7788.48, 7789.29) \\
O & (7771.53, 7772.27), (7773.75, 7774.57), (7775.08, 7775.75) \\
Sc & (4743.56, 4743.98), (4752.99, 4753.41), (5657.68, 5658.12), (5666.92, 5667.30), (5671.59, 5672.09), (5684.02, 5684.30), (5686.72, 5687.21), (5717.02, 5717.52), (5723.90, 5724.28), (6604.39, 6604.97) \\
Si & (5665.21, 5665.82), (5690.18, 5690.68), (5700.91, 5701.29), (5792.70, 5793.31) \\
Ti & (4719.32, 4719.60), (4757.96, 4758.28), (4759.07, 4759.48), (4764.40, 4764.82), (4778.06, 4778.43), (4781.56, 4781.93), (4797.84, 4798.12), (4798.35, 4798.63), (4801.80, 4802.21), (4820.11, 4820.66), (4849.04, 4849.41), (4865.28, 4865.83), (4873.88, 4874.20), (5689.25, 5689.80), (5716.25, 5716.80), (5720.27, 5720.65), (5739.24, 5739.68), (5866.02, 5866.79), (6598.89, 6599.53), (6716.52, 6716.90), (7852.19, 7853.01) \\
V & (4746.51, 4746.78), (4784.32, 4784.60), (4796.74, 4796.97), (4831.52, 4831.75), (4875.26, 4875.72), (5657.07, 5657.68), (5668.13, 5668.62), (5670.60, 5671.10), (5702.78, 5703.88), (5725.27, 5725.82), (5726.87, 5727.36), (5727.42, 5727.97), (5730.99, 5731.49), (5736.82, 5737.26), (5743.20, 5743.70), (6531.18, 6531.62) \\
Y & (4854.79, 4855.02), (4883.54, 4883.82), (5662.74, 5663.18), (5728.74, 5729.01) \\
Zn & (4721.99, 4722.27), (4810.36, 4810.63) \\
\enddata
\end{deluxetable}

\section{Action and azimuthal velocity uncertainty} \label{sec:kinematicerrors}
\begin{figure}
    \centering
    \includegraphics[width=0.45\textwidth]{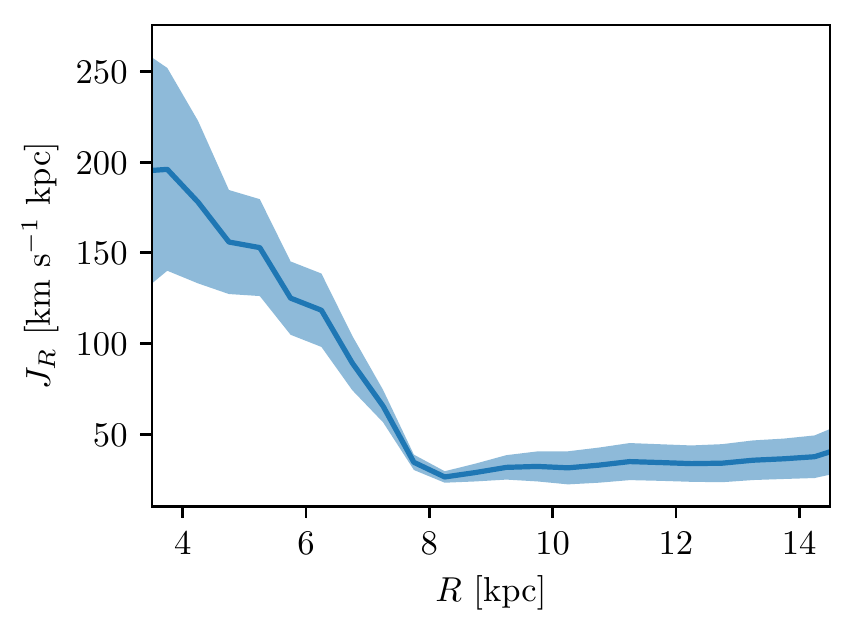}
    \includegraphics[width=0.45\textwidth]{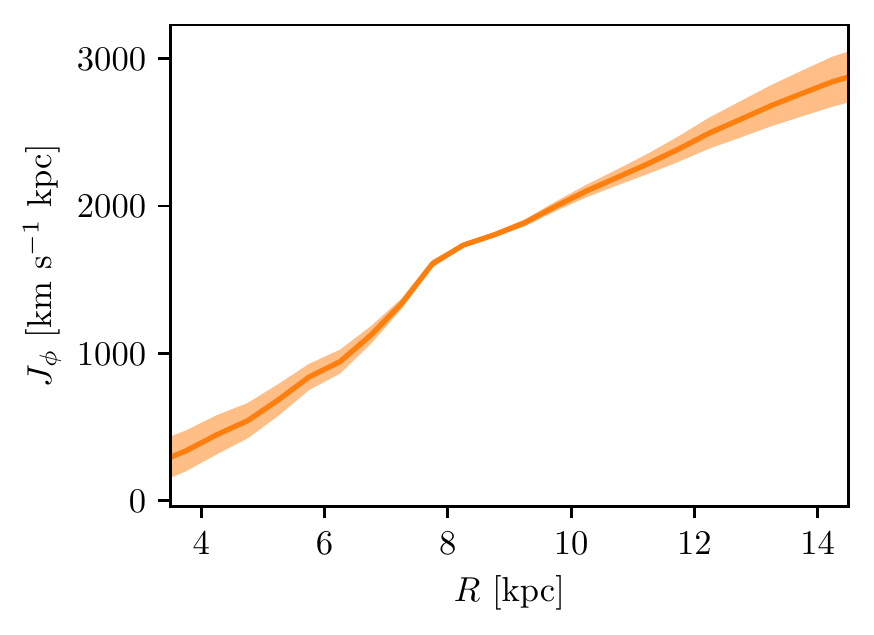}
    \includegraphics[width=0.45\textwidth]{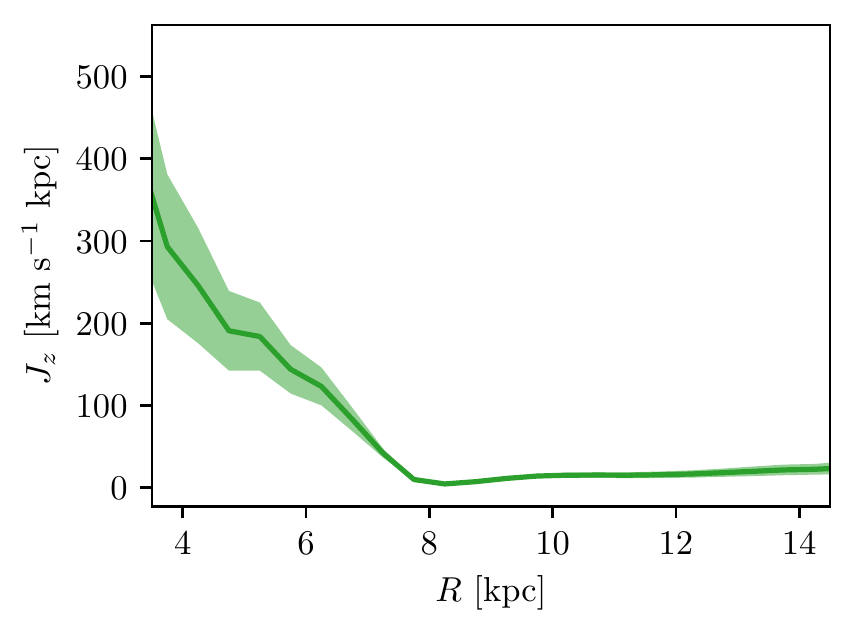}
    \includegraphics[width=0.45\textwidth]{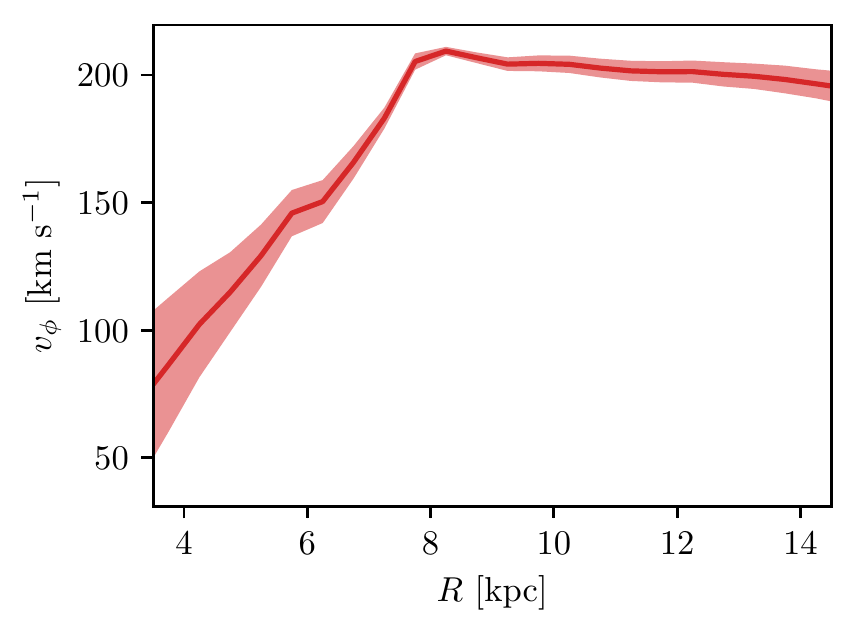}
    \caption{Median values and uncertainties of the three actions and azimuthal velocity as a function of Galactocentric radius, $R$. Note that the error bars here do not show scatter, but median uncertainty.}
    \label{fig:Jerr}
\end{figure}

Figure \ref{fig:Jerr} shows the median values and errors of the three actions and azimuthal velocity as a function of Galactocentric radius.
$J_R$ is particularly uncertain, and extremely so for stars interior to the sun.

\end{document}